\documentclass[manuscript]{aastex}

\usepackage{natbib, aas_macros}
\usepackage{amsmath}
\usepackage{color}

\usepackage{ulem}

\definecolor{MyDarkBlue}{rgb}{0,0.08,0.45}
\definecolor{MyDarkRed}{rgb}{0.8,0.1,0.08}
\definecolor{Red}{rgb}{1.0,0.0,0.2}
\definecolor{Blue}{rgb}{0,0.08,0.95}
\definecolor{LightGrey}{rgb}{0.7,0.7,0.7}

\begin{document}
\title{High-resolution calculation of the solar global convection with
the reduced speed of sound technique: II. Near surface shear layer with
the rotation}
\author{H. Hotta$^{1,2}$, M. Rempel$^1$, and T. Yokoyama$^2$}
\affil{
$^1$High Altitude Observatory, National Center for Atmospheric Research,
Boulder, CO, USA\\
$^2$Department of Earth and Planetary Science, University of Tokyo,
7-3-1 Hongo, Bunkyo-ku, Tokyo 113-0033, Japan
}
\email{ hotta@ucar.edu}
\begin{abstract}
We present a high-resolution, highly stratified numerical
simulation of rotating thermal convection in a spherical shell. Our
aim is to study in detail the processes that can maintain a near surface
shear layer (NSSL) as inferred from helioseismology. Using the reduced
speed of sound technique we can extend our global convection
simulation to $0.99\,R_{\odot}$ and include near the top of our domain
small-scale convection with short time scales that is only weakly
influenced by rotation. We find the formation of a NSSL preferentially
in high latitudes in the depth range $r=0.95-0.975R_\odot$.
The maintenance mechanisms are summarized as follows.
Convection under weak influence of rotation leads to Reynolds
stresses that transport angular momentum radially inward in all latitudes.
This leads to the formation of a strong poleward directed meridional flow and
a NSSL, which is balanced in the meridional plane by forces resulting from
the $\langle  v'_r v'_\theta\rangle$ correlation of turbulent velocities.
The origin of the required correlations depends to some degree on latitude.
In high latitudes a positive correlation $\langle  v'_rv'_\theta\rangle$ is induced 
in the NSSL by the poleward meridional flow whose amplitude increases with the radius,
while a negative correlation is generated by the Coriolis force in bulk of the convection 
zone. In low latitudes a positive correlation $\langle v'_rv'_\theta\rangle$ results from 
rotationally aligned convection cells (``banana cells''). The force caused by these Reynolds 
stresses is in balance with the Coriolis force in the NSSL.
\end{abstract}

\newpage
\section{Introduction}
\begin{figure}[htbp]
 \centering
 \includegraphics[width=15cm]{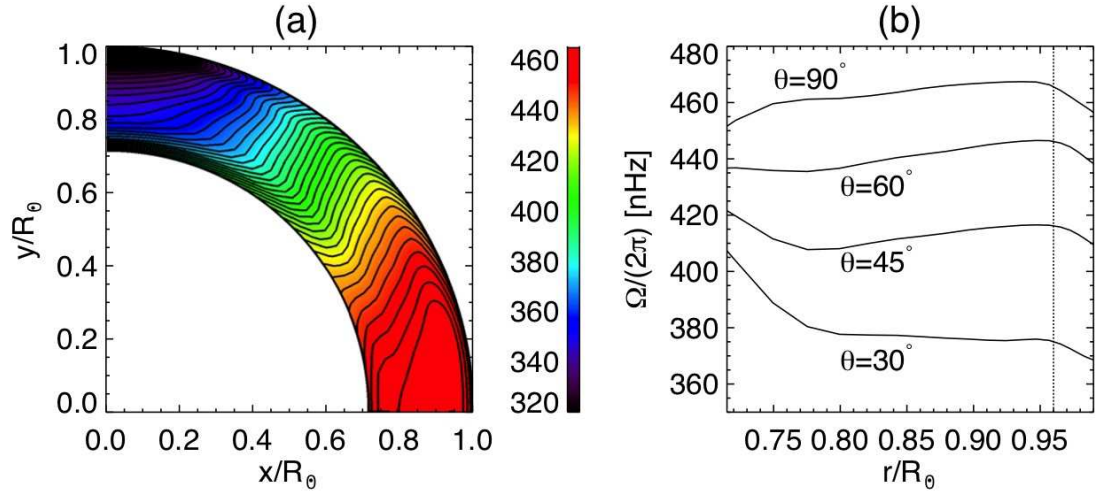}
 \caption{
 An inversion of the helioseismology from HMI data about the angular
 velocity ($\Omega/2\pi$) in the unit of nHz
 \citep{2011JPhCS.271a2061H} (a) on meridional plane and (b) along the
 selected colatitude. The dashed line in the panel b roughly shows the bottom of
 the NSSL.
 \label{dr_real}}
\end{figure}

The Sun is rotating differentially. The detailed distribution of the
angular velocity is revealed by helioseismology
\citep[][see
Fig. \ref{dr_real}]{2003ARA&A..41..599T,2009LRSP....6....1H,2011JPhCS.271a2061H}. The
three important
findings by helioseismology are the tachocline, the conical profile
in the middle of the convection zone, and the near surface shear layer
(NSSL). The physical mechanism underlying these features is discussed in the next
section. The three features show a significant deviation from the
expected Taylor-Proudman state where the angular velocity does not change along
the rotational axis. The strongest deviations from the Taylor-Proudman state are found
in the two boundary layers, the tachocline and the NSSL.
As shown in Fig. \ref{dr_real}, the thickness of the NSSL is about
0.04$R_\odot$, where $R_\odot$ is the solar radius. The difference of
the angular velocity ($\Omega/(2\pi)$) in the region is 10-20 nHz. The
variation in latitude is not significant in the NSSL (Fig. \ref{dr_real}b).
The existence of the NSSL 
was already inferred before the advent of helioseismology.
\cite{1975ApJ...199L..71F} pointed out that magnetic structures
rotate 5 \% faster than surrounding gas.
Then \cite{1984ApJ...283..373H} compared the rotation rate
estimated from the Doppler velocity measurement and the tracking of the
sunspots. It was found that the rotation rate of the sunspots is consistently
faster than the Doppler velocity. This was
interpreted as an indication that the sunspots are the
anchored in a faster-rotating deeper layer.\par
\subsection{Maintenance of the differential rotation}\label{mdr}
According to \cite{2011ApJ...743...79M}, the mean flows in the
convection zone are described by the two equations, which are the
gyroscopic pumping and the meridional force balance.
The discussion below is in the spherical geometry $(r,\theta,\phi)$.
The gyroscopic
pumping is derived from the zonal component of the equation of motion
with the anelastic approximation ($\nabla\cdot(\rho_0{\bf v})=0$)
as:
\begin{eqnarray}
\rho_0 \frac{\partial \langle\mathcal{L}\rangle}{\partial t} = 
 -\rho_0\langle{\bf v_\mathrm{m}}\rangle\cdot\nabla\mathcal{\langle
  L\rangle} + \mathcal{F}_\mathrm{R},
  \label{angular_momentum0}
\end{eqnarray}
where $\rho_0$, ${\bf v}_\mathrm{m}$, and
$\mathcal{L}=r\sin\theta u_\phi$ specify the background density,
the meridional flow, and the specific angular momentum. 
${\bf v}$ and ${\bf u}$ specify the fluid velocities at the rotating system and the
inertial reference system, respectively, i.e., 
${\bf u}={\bf v}+r\sin\theta\Omega_0{\bf e}_\phi$, where
$\Omega_0$ and ${\bf e}_\phi$ are the rotation rate of the
system and the zonal unit vector.
The bracket
$\langle\rangle$ indicates the average in time and zonal direction.
In this discussion, the magnetic field and the viscosity are neglected.
Then the term $\mathcal{F}_\mathrm{R}$ is expressed as:
\begin{eqnarray}
 \mathcal{F}_\mathrm{R} = -\nabla\cdot
  (\rho_0r\sin\theta \langle {\bf v}'_\mathrm{m}v'_\phi\rangle),
  \label{angular_momentum1}
\end{eqnarray}
where prime indicates the deviation from the axisymmetric temporally
averaged value, i.e.,
a value is divided as $Q=\langle Q\rangle+Q'$.
$\mathcal{F}_\mathrm{R}$ expresses the angular momentum transport by the Reynolds stress,
i.e., the non-linear coupling of the convective flow components. The gyroscopic
pumping equation
indicates that when the correlation of the convection flow is determined,
the mean
meridional flow is determined accordingly in the steady state
($\partial/\partial t=0$).
\par
The detailed derivation of the meridional force balance
is found in Appendix \ref{twb}.
\begin{eqnarray}
\frac{\partial \langle \omega_\phi\rangle}{\partial t} = 
[\langle\nabla\times({\bf v}\times{\bf \omega})\rangle]_\phi
+ 2r\sin\theta\Omega_0\frac{\partial \langle\Omega_1\rangle}{\partial z}
 + \frac{g}{\rho_0r}\left(\frac{\partial \rho}{\partial s}\right)_p
 \frac{\partial \langle s_1\rangle}{\partial \theta},
\label{eq_twb}
\end{eqnarray}
where ${\bf \omega}=\nabla\times{\bf v}$, $\Omega_1=v_\phi/(r\sin\theta)$,
$g$, and $s$ are the vorticity, the angular velocity, the
gravitational acceleration, and the entropy, respectively. The subscript
0 and 1 show the background and perturbed values, respectively. $z$
means the direction of the rotational axis.
The first term expresses the transport and the stretching which includes
both contributions of mean flow and turbulent flow (we call it
transport term).
The second term shows the Coriolis force on the meridional plane and the
third is the baroclinic term. 
\par
\begin{figure}[htbp]
 \centering
 \includegraphics[width=10cm]{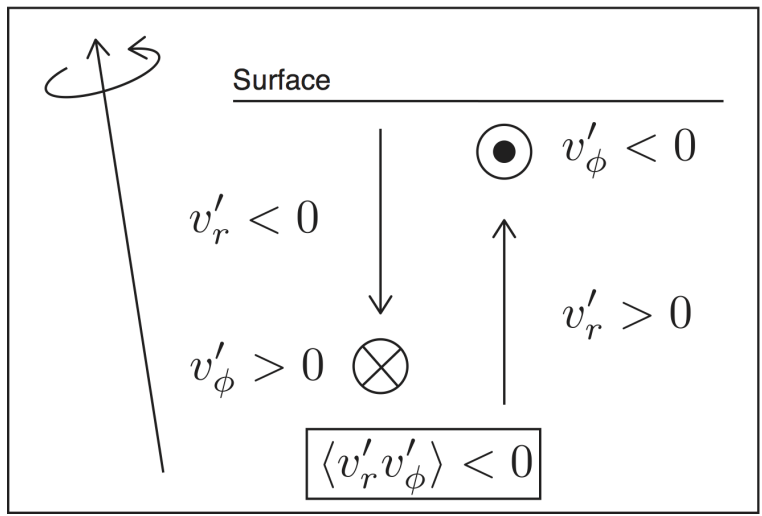}
 \caption{Schematic of the inward angular momentum transport by
 the radial velocity under the weak influence from the rotation.
 \label{radial_inward}}
\end{figure}

We discuss the NSSL with these two
equations. 
\cite{1975ApJ...199L..71F} suggest that when the convection
is not influenced much by the rotation, the radial
velocity in the thermal convection transports the angular momentum radially
inward (Fig. \ref{radial_inward}). When the influence from the rotation
is weak and the radial motion conserves the angular momentum,
the correlation 
$\langle v'_r v'_\phi\rangle$ is negative and transport angular momentum
radially inward. 
\cite{1975ApJ...199L..71F} argued that this is
the process for the generation and maintenance of the NSSL 
(see also
\cite{1979ApJ...229.1179G}).
There have been several attempts to reproduce the NSSL
based on this assumption
\citep{2002ApJ...581.1356D,2005ApJ...622.1320R,2007IAUS..239..457B,2013IAUS..294..417G}.
\par
\cite{2011ApJ...743...79M}, however, showed that the radially inward
angular momentum transport by the Reynolds stress is only a necessary
condition and that the force balance in the
meridional plane must be considered in addition. When the
transport term and the baroclinic term in eq. (\ref{eq_twb}) are neglected,
the meridional force balance equation becomes
\begin{eqnarray}
 \frac{\partial \langle\omega_\phi\rangle}{\partial t} = 2r\sin\theta
  \Omega_0\frac{\partial \langle\Omega_1\rangle}{\partial z}.
\end{eqnarray}
This means that when the radially inward angular momentum transport generates
the NSSL especially from mid to high latitudes, i.e.,
negative $\partial \langle\Omega_1\rangle/\partial z$, it creates an anti-clockwise
meridional flow.  This meridional flow continues to be accelerated and
transport the angular momentum, until $\partial \langle\Omega_1\rangle/\partial z$
becomes zero. 
Thus, in order to obtain a meridional force balance that breaks the
Taylor-Proudman constraint, other terms are necessary to compensate
the Coriolis force. For instance, it is thought that the baroclinic term
balances the Coriolis force within the bulk of the convection zone,
which may lead to the conical profile of the solar differential rotation
and the structure of the tachocline observed there
\citep{2005ApJ...622.1320R,2006ApJ...641..618M,2011ApJ...742...79B,2011ApJ...740...12H}.
\cite{2009MNRAS.395.2056B} obtained the solar-like differential rotation
with this idea and the assumption that isentropic and isorotational
surfaces coincide.
Regarding the NSSL, it
is unlikely that the baroclinic term is larger than that in the
middle of the convection zone. 
Even if so, the expected temperature would be 10 K at the surface,
which is not seen in observations \citep[e.g.][]{2008ApJ...673.1209R}.
The transport term could play
essential role in maintaining the NSSL.
In the near surface layer, the convection speed increases and the
spatial scale decreases. 
Indeed, it is expected that the ratio of the rotational period to the convective dynamical time
scale (i.e., the Rossby number Ro) should grow larger in the NSSL relative to the low Rossby
number ($\mathrm{Ro} < 1$) convection of the deep interior due to the decreasingly small overturning time
of convection near the surface.
Thus,
the reproduction of the NSSL in the numerical calculation requires a
wide range of spatial and temporal scales,
which must include giant cells
down to scales smaller than
supergranulation. 
Our previous study of non-rotating global convection was successful in capturing convective
scales smaller than supergranulation in the near surface layer using
the reduced speed of sound technique
\citep[][: hereafter Paper I]{2014ApJ...786...24H}.
In this study, we include the rotation to reproduce the NSSL in
the global convection calculation. The main focus of this paper is
to clarify the generation and maintenance mechanism of the NSSL in the view of 
the dynamical balance on the
meridional plane as well as the
angular momentum transport.
\section{Model}
\begin{figure}[htbp]
 \centering
 \includegraphics[width=15cm]{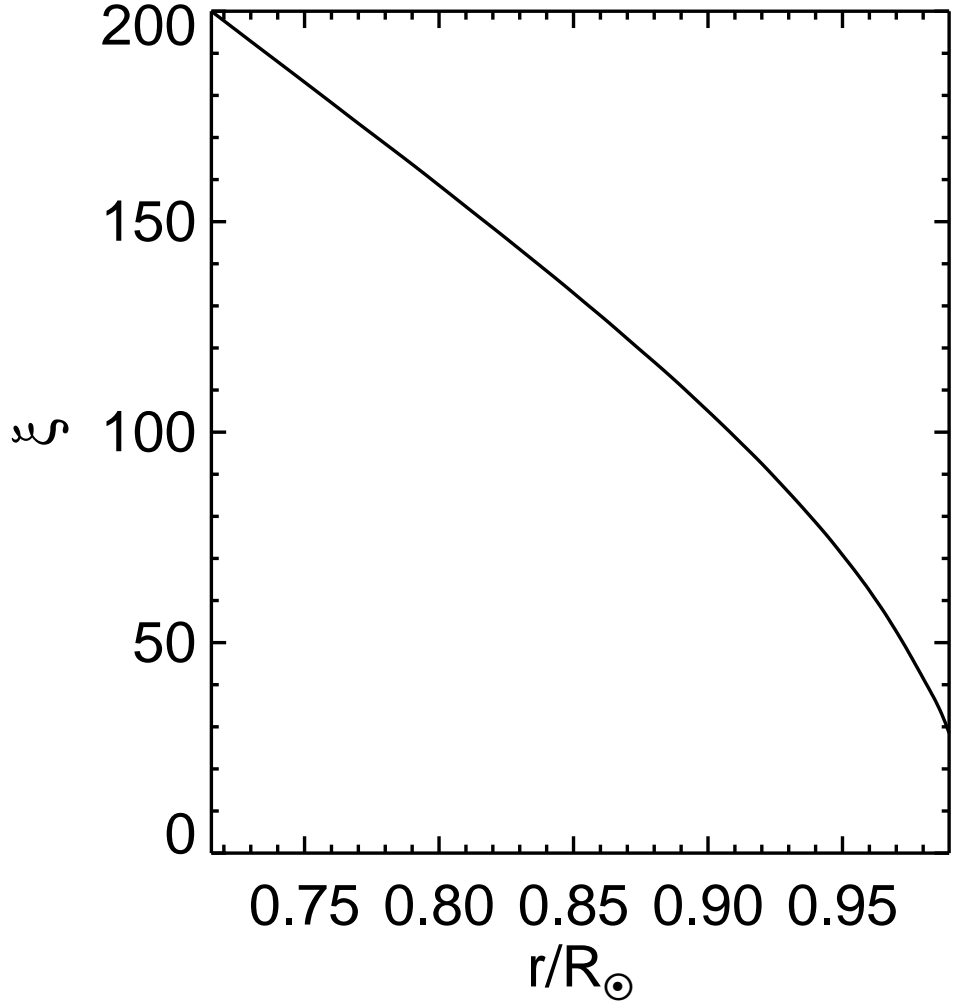}
 \caption{
The distribution of $\xi$.
 \label{xi}}
\end{figure}

We solve three-dimensional hydrodynamic equations in the spherical
geometry $(r,\theta,\phi)$:
\begin{eqnarray}
&&\frac{\partial}{\partial t}(\xi^2 \rho_1) =
 -\nabla\cdot\left(\rho {\bf
			     v}\right),\label{eqco}\\
&&\rho\frac{\partial {\bf v}}{\partial t} = -\rho({\bf
 v}\cdot\nabla){\bf v} - 
 \nabla p_1
 - \rho_1g{\bf e_r}+2\rho{\bf v}\times{\bf \Omega_0},\label{eqmo}\\
&& \rho T\frac{\partial s_1}{\partial t} = -\rho T({\bf v}\cdot\nabla)s_1
 +
\frac{1}{r^2}
\frac{d}{dr}
\left(r^2
  \kappa_\mathrm{r}\rho_0 c_\mathrm{p}\frac{dT_0}{dr} \right) +
 \Gamma,\label{eqen}\\
&& p_1 = \left(\frac{\partial p}{\partial \rho}\right)_s \rho_1 +
\left(\frac{\partial p}{\partial s}\right)_\rho s_1, \label{eos}
\end{eqnarray}
where $\rho=\rho_0+\xi^2\rho_1$.
The numerical model is similar to Paper I.
We adopt the new expression of reduced speed of sound technique
\citep[][see also Appendix \ref{rsst}]{2012A&A...539A..30H} and the
equation of state including the partial ionization effect for the Sun.
$\Gamma$ is the cooling term, which is effective only near
the surface.
We include the effect of rotation with a rate of $\Omega_0/(2\pi)=413\
\mathrm{nHz}$,
which is the solar rotation rate.
We adopt the
same artificial viscosity as \cite{2014ApJ...789..132R}. The details are
shown in Appendix \ref{art}. The same distribution of $\xi$ is used as
Paper I, which is defined as:
\begin{eqnarray}
 \xi(r) = \xi_0\frac{c_\mathrm{s}}{c_\mathrm{s}(r_\mathrm{min})},
\end{eqnarray}
where the adiabatic speed of sound is defined 
$c_\mathrm{s}=\sqrt{(\partial p/\partial \rho)_s}$ and $\xi_0=200$ is
adopted. Using this, the reduced speed of sound is 
$1.13\ \mathrm{km\ s^{-1}}$ at all depth.
The distribution of $\xi$ is shown in Fig. \ref{xi}.
The initial stratification is adiabatic $d s_0/d r=0$, and
a small perturbation is added to the entropy
in order to start convection.
The radiative
diffusivity is 18 times smaller than that calculated in the Model S
\citep{1996Sci...272.1286C};
thus the imposed luminosity is also 18 times smaller than the solar
luminosity.
When we use the
low viscosity
in combination with the solar
rotation rate and luminosity, the
polar region is accelerated
rather than the equator \citep{2013arXiv1305.6370F}.
There have been some systematic investigation on the relation between
the Rossby number and rotation profile
\citep{2011A&A...531A.162K,2011AN....332..897M,2014MNRAS.438L..76G}.
The formation of
the NSSL, however, requires the small-scale convection pattern, which
can be achieved only with low viscosity. Thus we use
the radiative diffusivity to decrease the Rossby number in the
convection zone until an acceleration of the equator is reproduced.
We implicitly assume that the numerically unresolved thermal convection transports
substantial energy in the real Sun.
We note that both high resolution and higher position of
the top boundary make it difficult to obtain the accelerated equator,
since both increase the Rossby number and are likely breaking
coherent rotationally aligned flows (``banana cells''). 
Thus a rather severe measure is required, i.e., 18 times
smaller luminosity, for
achieving a faster rotating equator in this study.
While our setup allows us to generate self-consistently a solar-like differential rotation and a near surface shear
layer, we have to be careful when applying our result to the Sun.
The resolution is
$384(N_r)\times648(N_\theta)\times1944(N_\phi)\times2$ in the Yin-Yang
grid, which is fairly high compared to other calculation
\citep[e.g.,][]{2008ApJ...673..557M} except for that in Paper I.
The top and bottom boundaries are at $0.99R_\odot$ and $0.715R_\odot$,
respectively.
Both boundaries are impenetrable and stress free, i.e.,
$v_r=\partial (v_\theta/r)/\partial r=\partial (v_\phi/r)/\partial r=0$.
A free boundary condition (zero gradient) is adopted for the density and entropy perturbation
($\partial \rho_1/\partial r=\partial s_1/\partial r=0$).
\section{Result}
\begin{figure}[htbp]
 \centering
 \includegraphics[width=15cm]{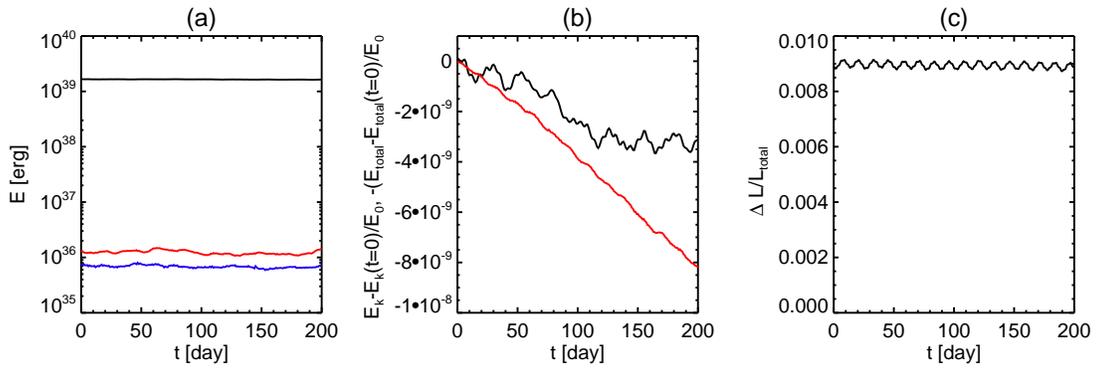}
 \caption{ (a) Temporal evolution of total kinetic energy of mean velocity.
 The blue, red, black lines show the total kinetic energy of
$\langle v_r\rangle$, $\langle v_\theta\rangle$, and
$\langle v_\phi\rangle$, respectively. 
(b) Temporal evolution of the difference of
 total kinetic energy including differential rotation, meridional flow
 and non-axisymmetric flow (black) and total energy including
 internal energy $\rho e_1 + \rho v^2/2$ (red) from the
 value at $t=0$ normalized by 
the background internal energy, i.e., the initial total energy. (c) Temporal
 evolution of the deviation of the angular momentum conservation using
 the ratio to the total angular momentum.
 \label{emean}}
\end{figure}

\begin{figure}[htbp]
 \centering
 \includegraphics[width=15cm]{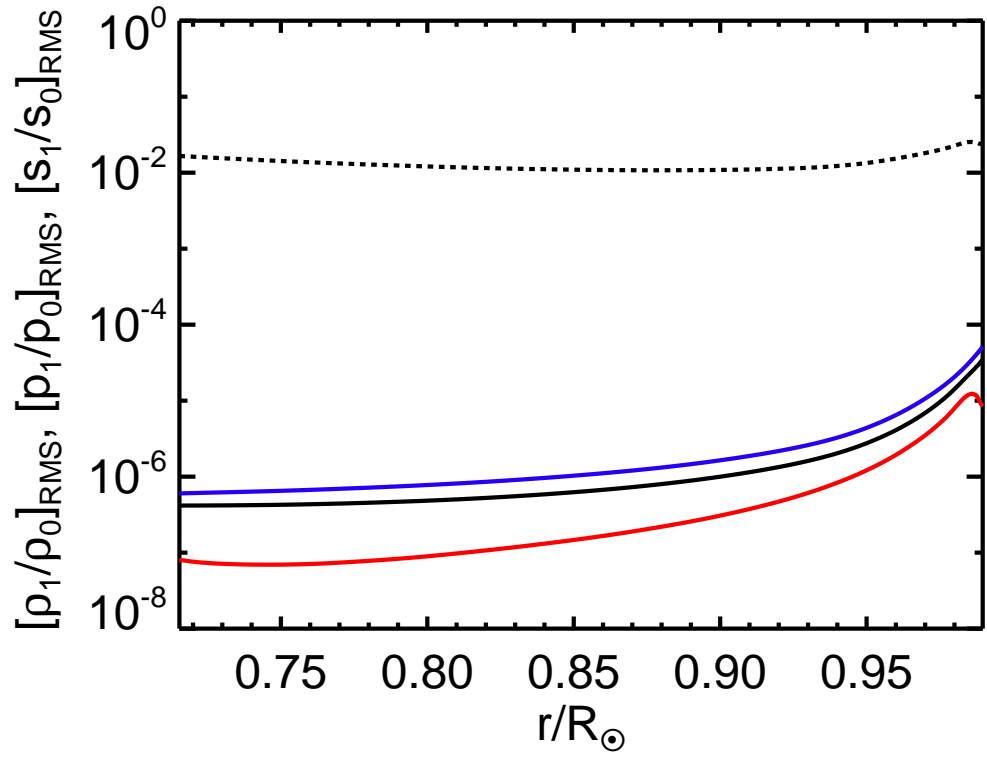}
 \caption{
The radial distribution of RMS values of 
$\rho_1/\rho_0$ (black), $p_1/p_0$ (blue) and $s_1/c_\mathrm{p}$ (red).
The dashed line shows RMS value of $\xi^2\rho_1/\rho_0$
 \label{etc}}
\end{figure}

\begin{figure}[htbp]
 \centering
 \includegraphics[width=16cm]{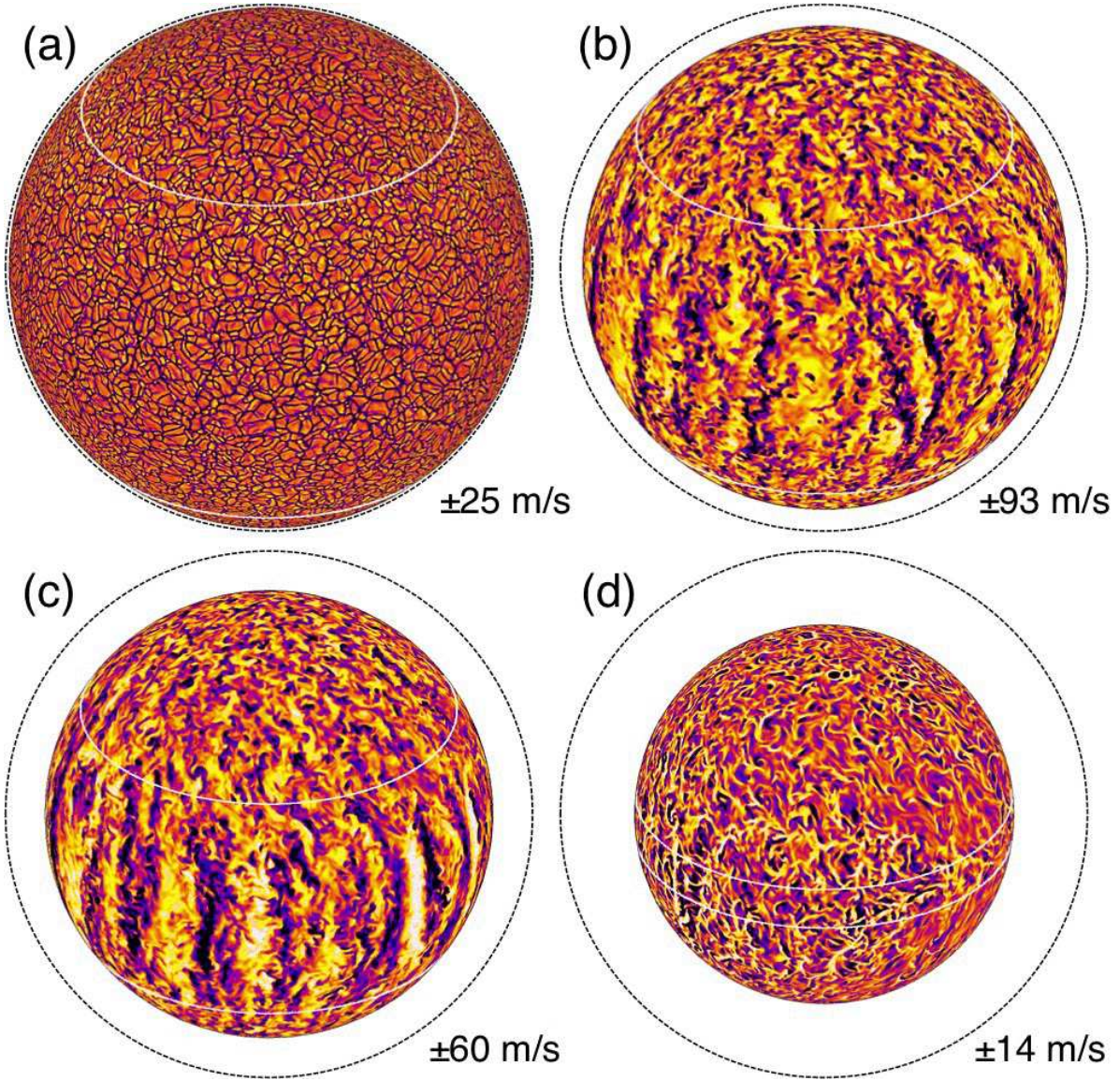}
 \caption{ Contour of the radial velocity $v_r$ at (a) $r=0.99R_\odot$
 (b) $r=0.92R_\odot$, (c) $r=0.85R_\odot$, (d) $r=0.72R_\odot$. The
 white lines show the tangential cylinder
$r\sin\theta=r_\mathrm{min}$,
 where $r_\mathrm{min}=0.715R_\odot$.
 \label{contour}}
\end{figure}

\begin{figure}
 \centering
 \includegraphics[width=16cm]{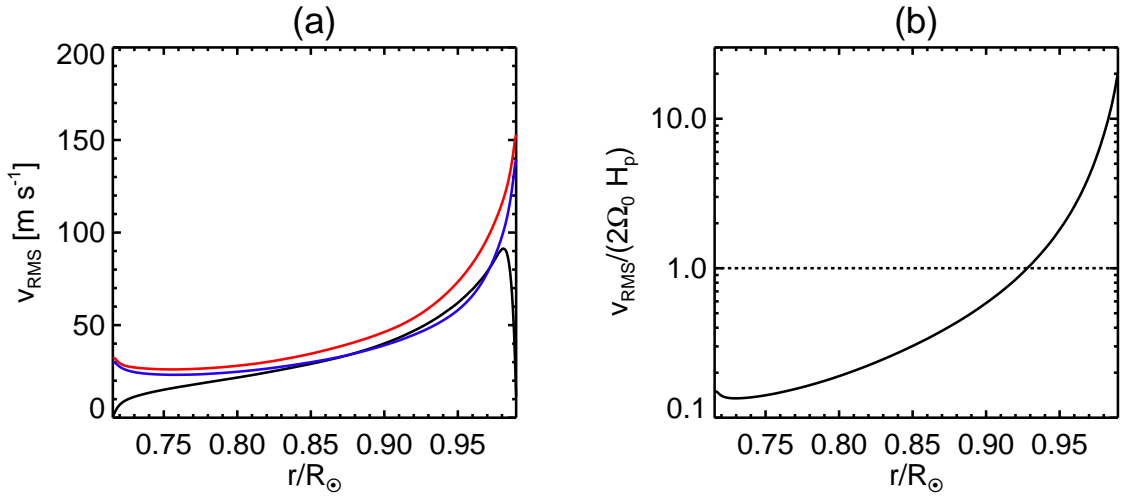}
\caption{
The radial profile of (a) the RMS velocity and
 (b) $v_\mathrm{RMS}/(2\Omega_0H_\mathrm{p})$. The black, blue and red
 lines show the radial ($v_r$), the latitudinal ($v_\theta$), and the
 zonal ($v_\phi$) values, respectively. The dashed line in the panel b
 indicates the values at unity.
\label{rms}}
\end{figure}

We use previously calculated data using higher artificial viscosity
and old expression of the RSST, i.e., using $\rho_0$ instead of $\rho$,
for 4500 days as a initial condition. Then we switch the expression to
current equations (eq. (\ref{eqco})-(\ref{eqen})) and reduce the artificial viscosity and calculate it for
200 days. Since the changes of differential rotation and
meridional flow are not so significant, 200 days
calculations are enough for the differential rotation and the meridional
flow to reach steady state, in which the time derivative of these
large-scale flows are small compared with the other terms.
In order to analyze the data, we continue the simulation for another 200 days.
Fig. \ref{emean}a shows the temporal evolution of the total energy of the
differential rotation ($\langle v_\phi\rangle$: black) and the
meridional flow ($\langle v_r\rangle$: blue and $\langle
v_\theta\rangle$: red).
The temporal evolution of total kinetic energy
(black) and total energy from $t=0$ are shown in Fig. \ref{emean}b. 
Since the plots of the energy (Fig. \ref{emean}b) indicate 
long term evolution, we also consider the influence of this evolution
with estimating $\partial \langle \mathcal{L}\rangle/\partial t$ and 
$\partial \langle \omega_\phi \rangle/\partial t$ in the following
analyses. 
Rather large time evolution is seen in the total energy
$\rho e_1 + \rho v^2/2$ (red line in Fig. \ref{emean}). We confirmed that this can be mostly explained
with the imbalance of
the energy flux between the bottom and the top boundary
caused by artificial viscosity on the entropy and the
radiative diffusion. This can be fixed in the future study.
This imbalance corresponds to about
3\% of the convective energy flux through the system and potentially
influences results on this level.
The conservation of the angular momentum
is reasonably confirmed in this period (Figs. \ref{emean}c).
RMS values of the density, the pressure and the entropy are shown in
Fig. \ref{etc}. These values are normalized by the background values
in order to show the validity of the linearized equation of state
(eq. (\ref{eos})). 
The dotted line shows the distribution of
$\xi^2[\rho_{1}/\rho_0]_\mathrm{RMS}$. 
Since  $\xi^2 \rho_1/\rho_0$ is 0.023 at maximum, it does not influence our
analyses by taking $\rho_0$ instead of $\rho=\rho_0+\xi^2 \rho_1$ 
and we mention this issue again in the following analyses.
We note that since we do not use $\xi^2\rho_1$ but
$\rho_1$ for the equation of state, the linearization of the equation of state is valid.
To increase the statistical validity, we average the north and south
hemispheres considering the symmetry.
Fig. \ref{contour}
shows the snapshot of the radial velocity $v_r$ at 
$t=200\ \mathrm{day}$ at selected depth, where $t=0$ is the start of the
analysis. (The corresponding movie is
available online.) The white lines show the location
of the tangential cylinder $r\sin\theta=r_\mathrm{min}$. We can
reproduce 10 Mm scale convection at $r=0.99R_\odot$ 
without any influence of the rotation in which
we cannot see any clear alignment of the convection pattern along the
rotational axis (the banana cell). At $r=0.92R_\odot$, the banana cell
like feature begins to appear and at $r=0.85R_\odot$, we can see clear
banana cell pattern. In addition, the banana cell
pattern is seen outside the tangential cylinder. This
dependence of the convection pattern on the depth is basically
determined by the Rossby number. 
Figs. \ref{rms}a and b show the radial profile of RMS velocity
and the Rossby number defined here by
$\mathrm{Ro}=v_\mathrm{RMS}/(2\Omega_0 H_p)$, respectively.
Three components of the RMS velocity
monotonically increase along with radius, whereas
$v_r$ monotonically decreases above $0.975R_\odot$ due to the top boundary condition.
The Mach number defined with RMS velocity and the
reduced speed of sound is 0.12 at maximum. This satisfies the criterion
obtained in \cite{2012A&A...539A..30H}.
This and the decrease in the
pressure scale height $H_p$ cause the significant increase of the Rossby
number around the surface. Especially above $r=0.93R_\odot$, the Rossby
number exceeds unity indicating weak rotational influence on the
convective flow.\par
\begin{figure}
 \centering
 \includegraphics[width=16cm]{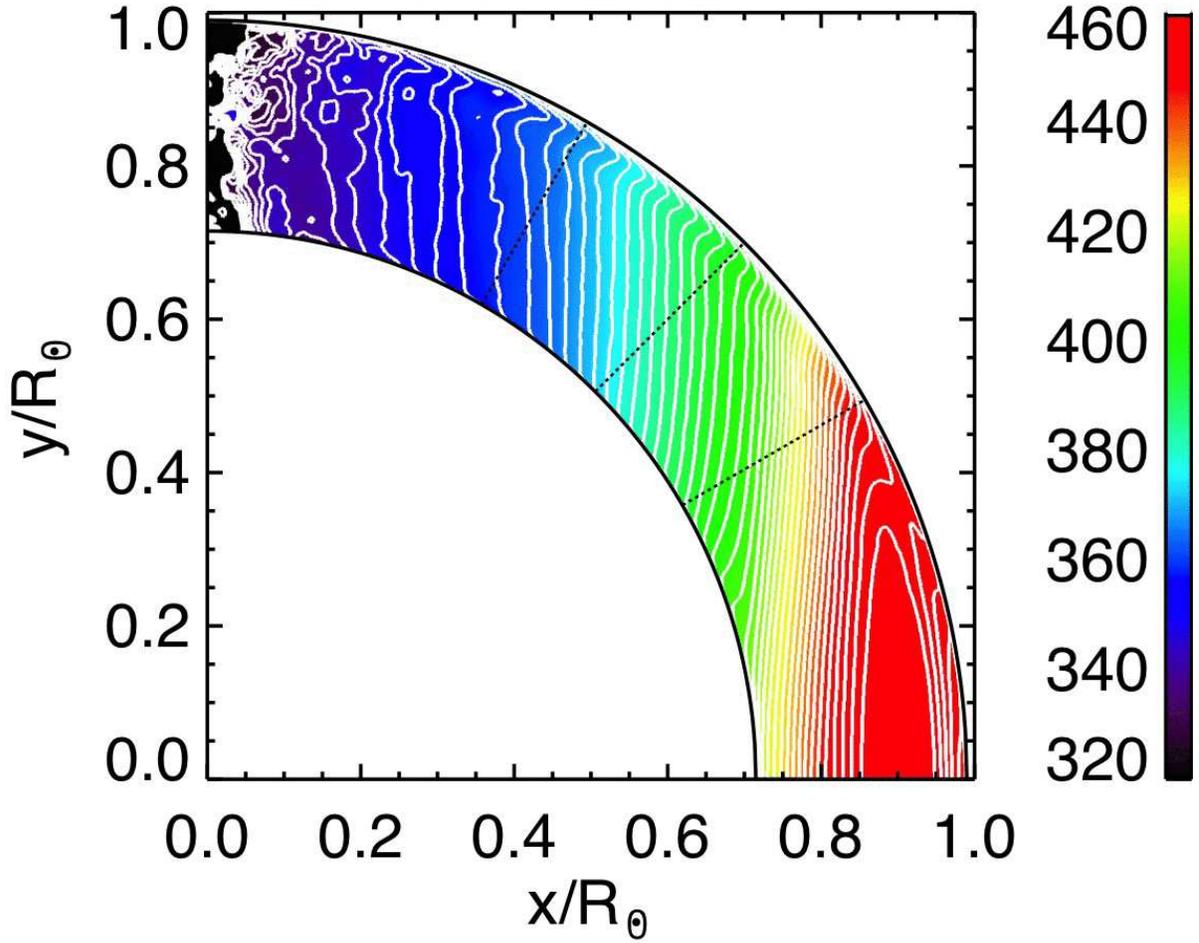}
\caption{The averaged angular velocity ($\langle\Omega\rangle/(2\pi)$)
 over 200 days in the unit of nHz. 
 The black dashed lines show the selected colatitude in
 Fig. \ref{dr_plot}.
\label{dr}}
\end{figure}
\begin{figure}
 \centering
 \includegraphics[width=16cm]{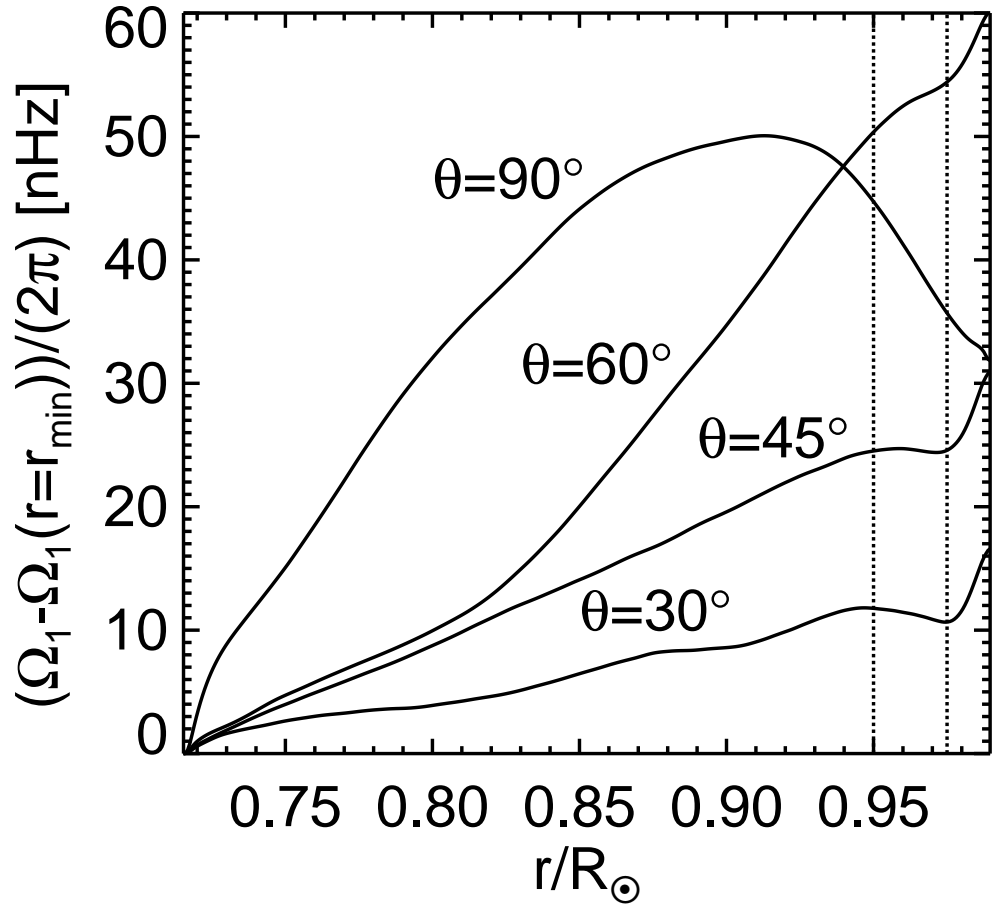}
\caption{
The radial profile of the angular velocity on the selected 
 colatitudes. 
The dotted lines shows $r=0.95R_\odot$ and $0.975R_\odot$,
 which is roughly the NSSL.
\label{dr_plot}}
\end{figure}

Fig. \ref{dr} shows the distribution of the angular velocity
($\langle\Omega\rangle/(2\pi)$), where $\Omega=\Omega_0 + \Omega_1$
and $\Omega_1=v_\phi/(r\sin\theta)$.
The NSSL's features are clearly seen especially in the 
low colatitude
($\theta>45\ \mathrm{degree}$) and high
colatitude ($\theta < 30\ \mathrm{degree}$).
We note that 
mid-colatitude is where poleward meridional flow is most
efficient at maintaining the Taylor-Proudman state, i.e., hardest to
maintain NSSL \cite[see also][]{2013IAUS..294..417G}.
In the convection zone at the
low to mid latitude, the
differential rotation is almost in the Taylor-Proudman state
($\partial \langle\Omega\rangle/\partial z\sim 0$). 
Note that the angular velocity has similar values to the solar one,
i.e., 460 nHz and 340 nHz at the equator and the polar regions,
respectively.
Fig. \ref{dr_plot} shows the radial
profile of the angular velocity at selected colatitude. At low colatitude
($\theta=30$ and $45\ \mathrm{degree}$), we can clearly see the decrease of the
angular velocity
from $r=0.95R_\odot$ to $0.975R_\odot$, which is the feature of the
NSSL. At the mid colatitude $\theta=60\ \mathrm{degrees}$ the tendency is
reversed. The angular velocity increases more steeply than that in
the deep convection zone.
At the high colatitude, the decrease from $r=0.92R_\odot$ to $0.99R_\odot$ is
seen.
The sign change of $\partial \Omega/\partial r$ above $0.975R_\odot$ is
related to the influence from the top boundary causing the RMS
value of $v_{r}$ to drop significantly.
\par
\begin{figure}[htbp]
 \centering
 \includegraphics[width=16cm]{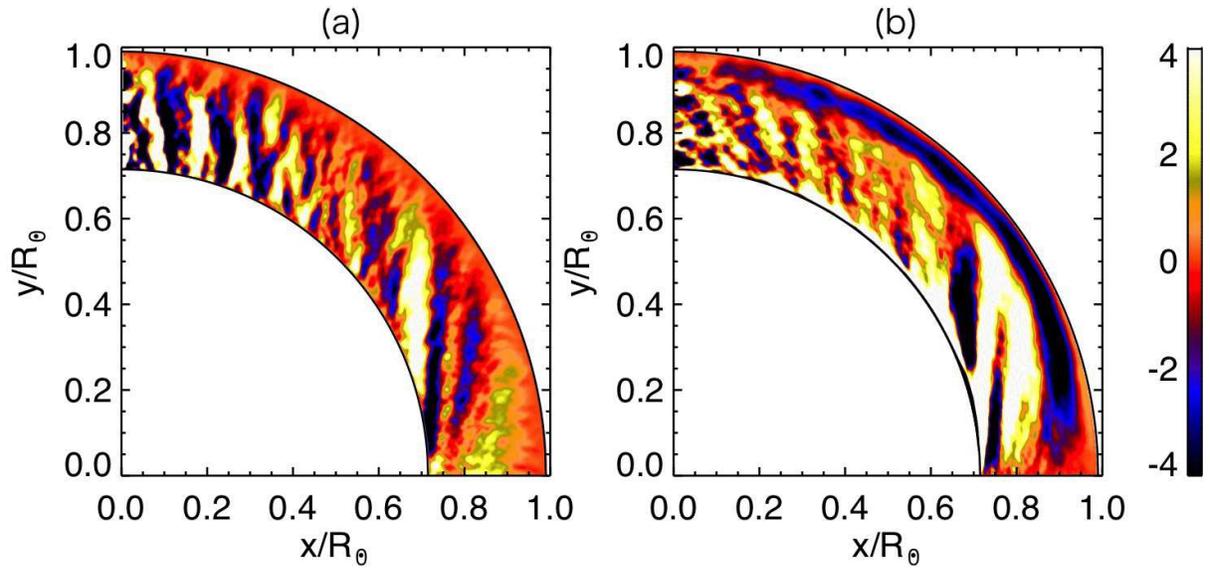}
 \caption{ The radial and latitudinal mass fluxes
 averaged in time and
 zonal direction over 200 days.
(a) $\rho_0\langle v_r\rangle$ and 
(b)$\rho_0\langle v_\theta\rangle$ in the unit of $\mathrm{g\ cm^{-2}\ s^{-1}}$.
 \label{mean_field}} 
\end{figure}

Fig. \ref{mean_field} shows the mean meridional flow.
Fig. \ref{mean_field}b clearly shows that
in the near surface area $(>0.9R_\odot)$, there is prominent poleward flow
which is caused by the radially inward angular momentum transport. 
An equatorward directed meridional flow is found
near the base of the convection zone and also in a thin layer round $0.85-0.9\,R_{\odot}$ below
$45\deg$ latitude.
 In the convection zone, the
multi-cell structure of the meridional flow is generated, which is
on qualitative level
similar to the recent finding by the local helioseismology
\citep{2013ApJ...774L..29Z}.
\par
\begin{figure}[htbp]
 \centering
 \includegraphics[width=16cm]{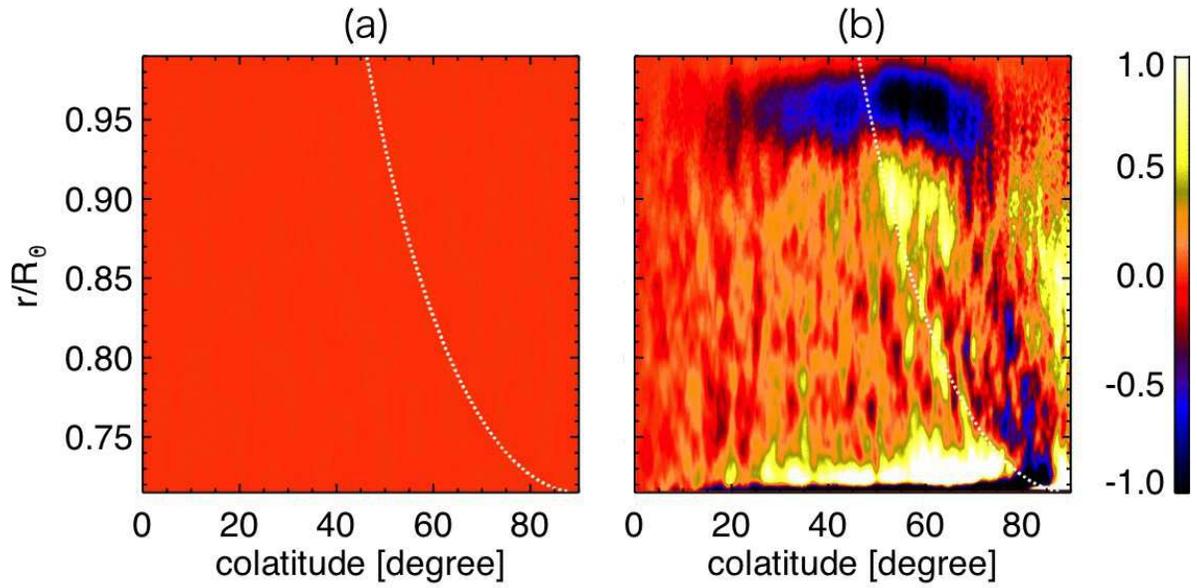}
 \caption{
 The values
 (a) $\langle \mathcal{L}\nabla\cdot(\rho{\bf v})\rangle$ and
 (b) $-\nabla\cdot(\rho_0\langle{\bf v'_\mathrm{m}} \mathcal{L'}\rangle)$, 
 in the unit of 
$10^6\ \mathrm{g\ cm^{-1}\ s^{-2}}$ are shown on the meridional
 plane. The white lines show the location of the tangential cylinder.
 \label{ang_divro}} 
\end{figure}
\begin{figure}[htbp]
 \centering
 \includegraphics[width=16cm]{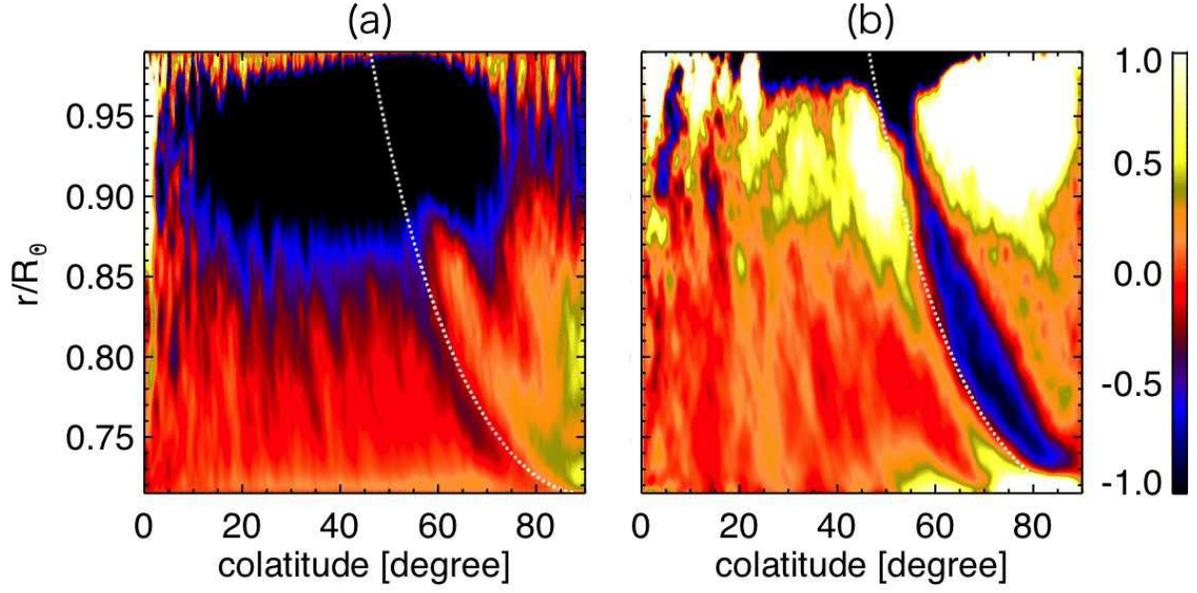}
 \caption{ The values (a) $\langle v'_r v'_\phi \rangle$, and
 (b) $\langle v'_\theta v'_\phi\rangle$ in the unit of $10^6\
 \mathrm{cm^2\ s^{-2}}$
 in the unit of $10^6\ \mathrm{g\ cm^{-1}\ s^{-2}}$
are shown on the meridional plane. The white lines show the location of
 the tangential cylinder.
 \label{ff}} 
\end{figure}

From our equation of motion, the balance equation for the specific
angular momentum is expressed as
\begin{eqnarray}
 \frac{\partial \langle \mathcal{L}\rangle}{\partial t}
= -\langle {\bf v_\mathrm{m}}\rangle \cdot \nabla \langle \mathcal{L}
\rangle
-\langle ({\bf v'_\mathrm{m}}\cdot\nabla) \mathcal{L}'\rangle
-r\sin\theta \left\langle \frac{\nabla \cdot {\bf
	      F_{v_\phi}}}{\rho}\right\rangle,
\label{conv_ang}
\end{eqnarray}
where the final term shows the artificial viscosity (see Appendix
\ref{art}). Figs. \ref{ang_balance}a, b, c, and d show 
$\rho_0\partial \langle \mathcal{L}\rangle/\partial t$,
$\rho_0\langle {\bf v_\mathrm{m}}\rangle\cdot\langle \mathcal{L}\rangle$, 
$-\rho_0\langle ({\bf v'_\mathrm{m}}\cdot\nabla)\mathcal{L}'\rangle$
and (d) $-\rho_0r\sin\theta\left\langle \nabla\cdot {\bf
 F_{v_\phi}}/\rho \right\rangle$, respectively. The background density
 $\rho_0$ is multiplied to see the balance in the convection zone and
 the near surface area simultaneously. The balance between angular
 momentum transports by the mean flow (panel b) and turbulence (panel c)
 is fairly good.
 Since the term $\rho_0\partial\langle\mathcal{L}\rangle/\partial t$ is
 small compared with other term, the distribution of angular momentum is almost in
 steady state. The L2 norm of 
$\partial \langle \mathcal{L}\rangle/\partial t$ is 0.04\% of the sum of
L2 norm of the terms in the right hand of eq. (\ref{conv_ang}).
The effect of the artificial viscosity is seen only
 around the bottom boundary. 
This would be caused by the thin fast down
 flow crashing to the bottom wall boundary.
In order to have a discussion with the
 Reynolds stress, we consider the relation
\begin{eqnarray}
 \rho ({\bf v}\cdot\nabla)\mathcal{L} = 
\mathcal{L}\nabla\cdot(\rho {\bf v}) + \nabla\cdot(\rho {\bf v}\mathcal{L}).
\end{eqnarray}
Figs. \ref{ang_divro}a and b show 
$\langle \mathcal{L}\nabla\cdot(\rho {\bf v})\rangle$ and
$-\nabla\cdot(\rho_0\langle{\bf v'_\mathrm{m}}\mathcal{L'}\rangle)$,
respectively. The contribution related to $\nabla\cdot(\rho{\bf v})$ is very
small and the values
$-\rho_0\langle ({\bf v'_\mathrm{m}}\cdot\nabla)\mathcal{L}'\rangle$ and 
$-\nabla\cdot(\rho_0\langle{\bf v'_\mathrm{m}}\mathcal{L}'\rangle)$ are
almost equivalent. Thus we can use the Reynolds stress, i.e., the
correlation of velocities, to understand the balance of angular
momentum. 
We confirmed the relations of
$\xi^2\langle \rho'{\bf v'_\mathrm{m}}
\mathcal{L}'\rangle \ll\rho_0\langle{\bf
v'_\mathrm{m}}\mathcal{L}'\rangle$ 
and $\xi^2\langle\rho'{\bf v'_\mathrm{m}}\rangle \ll \rho_0 \langle
\mathrm{v_\mathrm{m}}\rangle$, 
where $\rho' = \rho_1-\langle \rho_1\rangle$.
Figs. \ref{ff}a and b show the correlations between the velocities, that
is $\langle v'_rv'_\phi\rangle$ and $\langle v'_\theta
v'_\phi\rangle$. We note that these correlations are not normalized by
the RMS velocity (different from the definition in Paper I).
The negative correlation of $\langle v'_rv'_\phi\rangle$
which is speculated by Fig. \ref{radial_inward} is reproduced, which
causes the radially inward
angular momentum transport. This negative correlation is not confined to
the NSSL. 
In contrast, at low-latitudes, a positive correlation of
$\langle v'_\theta v'_\phi \rangle$ is realized, and is likely due to
the banana-cell-like features \citep{2005LRSP....2....1M}.
\par
\begin{figure}[htbp]
 \centering
 \includegraphics[width=16cm]{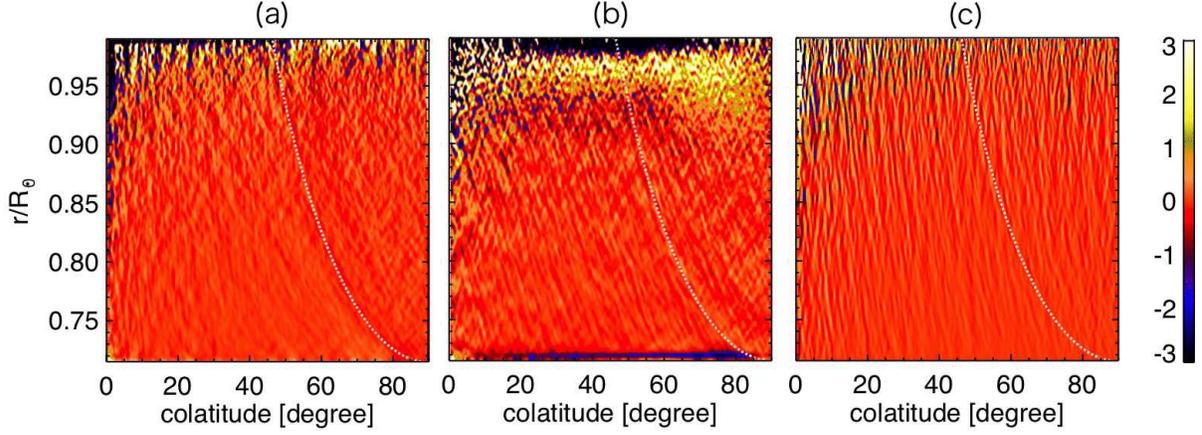}
 \caption{ The values (a) $\mathcal{C}'_\mathrm{d}$
(b) $\mathcal{C}'_\theta$, and
(c) $\mathcal{C}'_r$ in the unit of $10^{-12}\ \mathrm{s^{-2}}$.
are shown on the meridional plane. The white lines show the location of
 the tangential cylinder.
 \label{divide}} 
\end{figure}

As introduced in \S \ref{mdr}, the discussion regarding the
meridional force balance is required to understand the maintenance
mechanism of the NSSL
in addition to the angular momentum transport shown in Fig. \ref{ff}.
We discuss the dynamical balance by using the correlation of velocities.
Thus we check the relation of 
\begin{eqnarray}
 \nabla\times(\langle{\bf v}\times{\bf \omega}\rangle) = -\nabla\times
  \left[
   \left\langle\frac{1}{\rho}(\langle\nabla\cdot(\rho{\bf vv})
   -{\bf v}\nabla\cdot(\rho{\bf v}))\right\rangle
  \right].
\end{eqnarray}
Figs. \ref{compare}a, and c show the first term of the left hand side and
the second term of the right hand side,
respectively. Fig. \ref{compare}b shows the value
$\mathcal{C}=-\nabla\times[\nabla\cdot(\rho_0\langle{\bf vv}\rangle)/\rho_0]$. 
Figs. \ref{compare}a and b are almost equivalent and the
value related to $\nabla\cdot(\rho{\bf v})$ is negligible. Thus, we use
$\mathcal{C}$ as the transport term instead of
$\nabla\times(\langle{\bf v}\times{\bf \omega}\rangle)$.
We divide the contribution of the dynamical balance on the meridional
plane as:
\begin{eqnarray}
 \mathcal{W-T} = \mathcal{B}+\tilde{\mathcal{C}}+\mathcal{C}'+\mathcal{V},
\end{eqnarray}
where
\begin{eqnarray}
&& \mathcal{W} = \frac{\partial \langle \omega_\phi \rangle}{\partial
 t}, \\
 && \mathcal{T} = r\sin\theta\frac{\partial \langle \Omega\rangle^2}{\partial z},\\
&& \mathcal{B} = 
 -\left[
\nabla\times
\left(
\frac{\nabla p_1 + \rho_1 g{\bf e_r}}{\rho}
\right)
\right]_\phi.
\end{eqnarray}
The term $\mathcal{T}$ is caused by the Coriolis force on the meridional
plane, which contributes to the balance when the differential rotation
deviates from the Taylor-Proudman state
($\partial \langle\Omega\rangle/\partial z\neq 0$). The term 
$\mathcal{B}$ is caused by the pressure gradient and the buoyancy
(baroclinic term) and requires a
latitudinal entropy gradient to be present. The detailed
form of $\mathcal{C}'$ and $\tilde{\mathcal{C}}$ are found in the
Appendix \ref{twb}. These two are caused by the momentum transport within
the meridional plane. $\mathcal{C}'$ and $\tilde{\mathcal{C}}$ are
contribution by
the mean meridional flow
($\langle v_r \rangle$ and $\langle v_\theta \rangle$) and the
non-axisymmetric flow ($v'_r$ and $v'_\theta$), respectively.
$\mathcal{V}$ is contribution from the artificial viscosity
(see Appendix \ref{art} and \ref{twb}).
Fig. \ref{term} shows the distribution of (a) $\mathcal{W}$, (b)
$-\mathcal{T}$, (c) $\mathcal{B}$,
(d) $\tilde{\mathcal{C}}$, (e)
$\mathcal{C'}$, and (f) $\mathcal{V}$.
According to the distribution of the $-\mathcal{T}$, we divide the
meridional plane to four regions
(I, II, III, and IV as shown in Fig. \ref{term}b). Region I is the
maintained by the latitudinal entropy gradient $\mathcal{B}$ from the middle to the
bottom of the convection zone. 
In the other regions (II, III, and IV), the deviation
from the Taylor-Proudman state cannot be explained by the entropy
gradient alone. 
The contributions from time evolution $\mathcal{W}$ (panel a), mean flow
$\tilde{\mathcal{C}}$ (panel d), and artificial viscosity $\mathcal{V}$ (panel f)
have negligible role even in the NSSL.
Then we see that the contribution
from the non-axisymmetric flow ($\mathcal{C}'$: Fig. \ref{term}d) is almost
totally compensates the term $-\mathcal{T}$ at the regions II, III, and
IV. 
To investigate the origin of the distribution of
$\mathcal{C}'$, which can maintain the NSSL, we divide the term
$\mathcal{C}'$ to three as
$\mathcal{C}'=\mathcal{C}'_\mathrm{d}+\mathcal{C}'_\theta+\mathcal{C}'_r$.
The detailed forms of them are found in the Appendix \ref{twb}. The term
$\mathcal{C}'_\mathrm{d}$ is caused by the diagonal momentum flux
$F'_{rr}$, $F'_{\theta\theta}$, and $F'_{\phi\phi}$, where 
$F'_{ij}=\rho_0\langle v'_i v'_j\rangle$ (see Appendix \ref{twb}). The terms
$\mathcal{C}'_\theta$ and $\mathcal{C}'_r$ are caused by the
non-diagonal momentum flux $F'_{r\theta}$. The difference of these two
terms is explained as:
The term $\mathcal{C}'_\theta$ ($\mathcal{C}'_r$) is
caused by the transport of the latitudinal momentum $\rho_0v'_\theta$
(radial momentum $\rho_0v'_r$) in the radial (latitudinal) direction.
We note that $\mathcal{C}'_r$ and $\mathcal{C}'_\theta$ can act as
turbulent diffusivity on meridional flow.
Fig. \ref{divide} shows the distribution of (a)
$\mathcal{C}'_\mathrm{d}$, (b) $\mathcal{C}'_\theta$, and 
(c) $\mathcal{C}'_r$. The diagonal term $\mathcal{C}'_\mathrm{d}$ has
 contribution to some degree and the contribution from the term
 $\mathcal{C}'_r$ is negligible. The essential contribution is by the
 term $\mathcal{C}'_\theta$, i.e., the transport of the latitudinal
 momentum in the radial direction. \par
\begin{figure}[htbp]
 \centering
 \includegraphics[width=16cm]{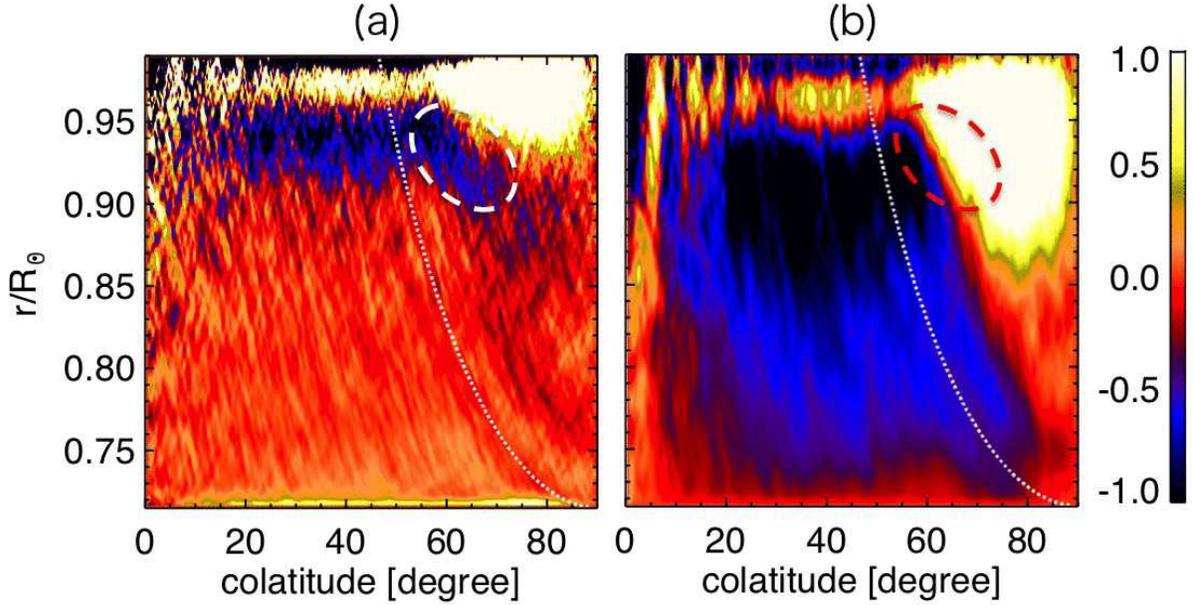}
 \caption{ The values (a) $D_{\theta\mathrm{(n)}}$ in the unit of 
 $10^{-3}\ \mathrm{cm\ s^{-2}}$ and 
 (b) $\langle v'_r v'_\theta\rangle$ in the unit of 
 $10^6\ \mathrm{cm^2\ s^{-2}}$
are shown on the meridional plane. The white lines show the location of
 the tangential cylinder. 
Dashed circle shows the boundary of the effective and ineffective area
 of the banana cell.
 \label{vxvy}} 
\end{figure}

Next we investigate the origin of $\mathcal{C}'_\theta$ by estimating the quantity $D'_{\theta(n)}$ , which is latitudinal
force arising from momentum transport (see Appendix \ref{twb} for a more complete definition of $D$).
This force is defined as
\begin{eqnarray}
 D'_{\theta(n)} = -\frac{1}{\rho_0}
\left[
 \frac{1}{r^2}\frac{\partial }{\partial r}(r^2F'_{r\theta})-\frac{F'_{\theta r}}{r}
\right],
\end{eqnarray}
which it is related to $\mathcal{C}'_\theta$
\begin{eqnarray}
 \mathcal{C'}_\theta=\frac{1}{r}\frac{\partial}{\partial  r}
\left(rD'_{\theta(n)}\right),
\end{eqnarray}
and where the subscript $\theta(n)$ refers to the inertial force in the latitudinal direction arising from
the non-diagonal Reynolds stress 
$F'_{r\theta} = \langle v'_r v'_\theta\rangle$. 
Fig. \ref{vxvy} shows (a) $D'_{\theta(n)}$ and (b) $\langle v'_r v'_\theta\rangle$,
where
in (a) it is evident that the direction of the inertial force is equatorward (poleward) at the top
(bottom) of the NSSL at high latitudes (i.e., Region II). In this region, the inertial force tends to
balance the Coriolis force. The origin of this inertial force is the
Reynolds stress $\langle v'_r v'_\theta\rangle$, as can be
deduced from the correlation of Figs. \ref{vxvy}b. In the high-latitude NSSL, the positive correlation
$\langle v'_r v'_\theta\rangle>0$ leads to the upward transport of
latitudinal momentum. In contrast, in the high-latitude
deep convection zone, the correlation is negative. This arrangement of
momentum flux increases
(decreases) the latitudinal momentum in the upper (lower) part of the NSSL (Fig. \ref{vxvy}a). These
correlations are the essential ingredients that maintain the meridional flow within the NSSL at
high latitudes.
\par
The following discussion is centered around the origin of velocity
correlations generated from a combination of rotation and large scale
shear. 
We retain the dominant terms that can generate a positive or negative correlation as:
\begin{eqnarray}
 \frac{\partial v'_r}{\partial t} &=& -\frac{v'_\theta}{r}\frac{\partial \langle
  v_r \rangle}{\partial \theta} + 2v'_\phi\langle\Omega\rangle\sin\theta+[...],\label{core1}\\
 \frac{\partial v'_\theta}{\partial t} &=& -v'_r\frac{\partial \langle
  v_\theta \rangle}{\partial r} + 2v'_\phi\langle\Omega\rangle\cos\theta+[...],\label{core2}\\
 \frac{\partial v'_\phi}{\partial t} &=&
-2v'_r\langle\Omega\rangle\sin\theta
-2v'_\theta\langle\Omega\rangle\cos\theta+[...].
\label{core3}
\end{eqnarray}
The sign of velocity correlation significantly depends on whether
$v'_r$ and $v'_\theta$ is generated by $v'_\phi$ (Situation 1) or
$v'_\phi$ is generated by $v'_r$ and $v'_\theta$ (Situation 2). 
The signs of $\langle v'_r v'_\phi\rangle$ and $\langle v'_\theta
v'_\phi\rangle$ are the direct consequence of these situations.
When the Situation 1 is achieved, a positive
correlations 
($\langle v'_r v'_\phi\rangle$ and $\langle v'_\theta v'_\phi\rangle$) are
generated through eqs. (\ref{core1}) and (\ref{core2}). On the other
hand, under the Situation 2, a negative correlations are generated
through eq. (\ref{core3}).
Figs. \ref{ff}a and b indicate that the Situation 1 requires both the
low Rossby number and the banana cell, i.e., deeper layer and outside
the tangential cylinder, since the
positive correlations ($\langle v'_r v'_\phi\rangle$ and 
$\langle v'_\theta v'_\phi\rangle$) are especially seen there.
Outside the tangential cylinder with low Rossby number, the zonal flow
$v'_\phi$ is dominant
due to coherent banana cell structure with weak influence from the
bottom boundary
\citep{1979ApJ...231..284G,2000ApJ...532..593M,2005LRSP....2....1M,2011ApJ...742...79B}. Thus
$v'_\phi$ generates $v'_r$ and $v'_\theta$ there.
In contrast to the Situation 1, the Situation 2 is realized even in high
Rossby number. The role of the meridional flow, however, becomes large
in a high Rossby number situation (see the following discussion).
\par
\begin{figure}[htbp]
 \centering
 \includegraphics[width=16cm]{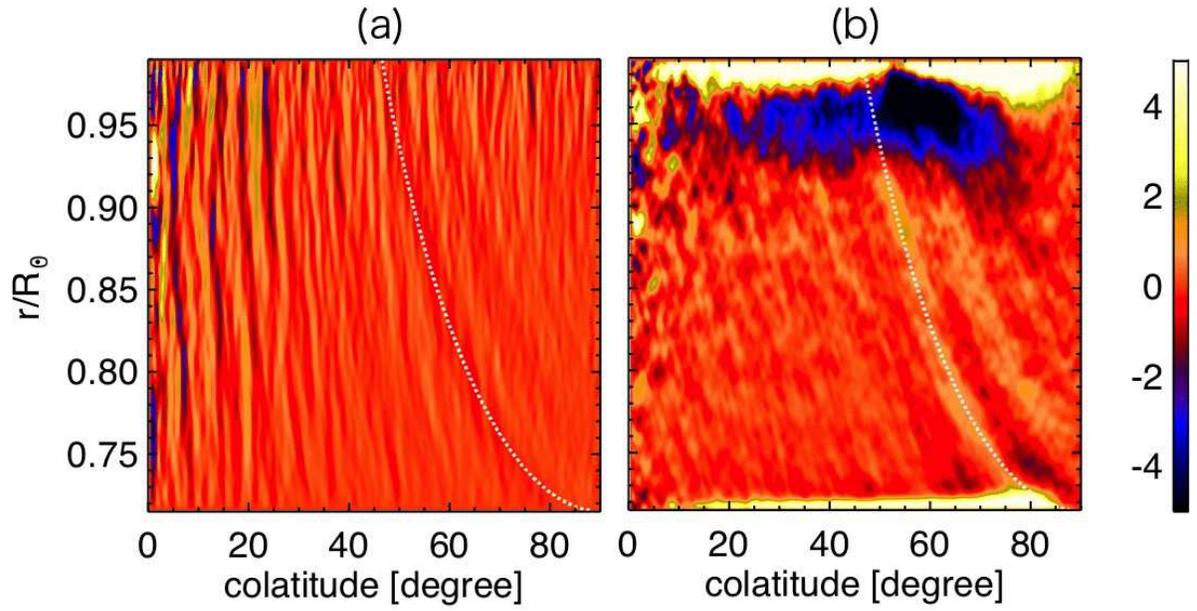}
 \caption{
 The values (a) $\partial \langle v_r \rangle/(r\partial \theta)$, and
 (b) $\partial \langle v_\theta \rangle/\partial r$ in the unit of
 $10^{-7}\ \mathrm{s^{-1}}$
are shown on the meridional plane. The white lines show the location of
 the tangential cylinder. 
 \label{mean_deri}} 
\end{figure}

Before discussing the origin of the correlation $\langle v'_r
v'_\theta\rangle$
in the high-latitude NSSL, we first describe the feature found in low latitudes.
From the high- to the mid-latitudes we find a negative correlation
in the near surface layer, while a positive
correlation is generated in the
lower latitude from the surface to the middle of the convection zone.
This positive correlation is generated by the banana
cells. 
When both the radial and latitudinal velocities are
generated by the Coriolis force, the correlation
$\langle v'_rv'_\theta\rangle$ can be positive (see also
eq. (\ref{core1}) and (\ref{core2})).
In the NSSL, however,
the Rossby number is large and banana cells do not exist.
This means that the positive correlation $\langle v'_rv'_\theta\rangle$
in the high latitude NSSL
is generated by different mechanism(s).
The first term in each of the eqs. (\ref{core1}) and (\ref{core2}) is
that due to the mean meridional flow that is the most important element
in this discussion.
In this discussion, we focus on the correlation between $v'_r$ and
$v'_\theta$. When the typical time scale is estimated
as $\tau=H_p/v_\mathrm{RMS}$, we obtain the relation
\begin{eqnarray}
 v'_\phi \sim  -2\tau v'_r\langle\Omega\rangle\sin\theta-2\tau v'_\theta\langle\Omega\rangle\cos\theta,
\end{eqnarray}
from eq. (\ref{core3}). 
Note that we can use this transformation, since the region is
inside the tangential cylinder
where no banana cell exists and $v'_\phi$ is generated by $v'_r$ and
$v'_\theta$ (see the discussion in previous paragraph).
We substitute this relation to
eqs. (\ref{core1}) and (\ref{core2}) and only
retain the terms that can generate the nonzero correlation between
$v'_r$ and $v'_\theta$:
\begin{eqnarray}
 \frac{\partial v'_r}{\partial t} = [...]-\frac{v'_\theta}{r}\frac{\partial \langle
  v_r \rangle}{\partial \theta} - 2v'_\theta\tau\langle\Omega\rangle^2\sin(2\theta),\\
 \frac{\partial v'_\theta}{\partial t} = [...]-v'_r\frac{\partial \langle
  v_\theta \rangle}{\partial r} - 2v'_r\tau\langle\Omega\rangle^2\sin(2\theta),
\end{eqnarray}
This means that the terms from the Coriolis force
(i.e., the last term in each equation)
 generates a negative
correlation between $v'_r$ and $v'_\theta$.
This is expected since a strong Coriolis force leads to fluid motions
preferentially aligned with the axis of rotation.
The sign of the correlation by the mean
flow depends on the sign of 
$\partial \langle v_r\rangle/(r\partial \theta) $ and
$\partial \langle v_\theta\rangle/\partial r$. Fig. \ref{mean_deri}
shows the distribution of 
(a)
$\partial \langle v_r\rangle/(r\partial\theta) $, and
(b) $\partial \langle v_\theta\rangle/\partial r$. It is clear that the
contribution from the term related to 
$\partial \langle v_r\rangle/(r\partial \theta)$ is small compared with
the term of $\partial \langle v_\theta \rangle/\partial r$. 
Interestingly we find a negative value of
$\partial \langle v_\theta \rangle/\partial r$ in region II and a
positive value in region IV. Only when 
$\partial \langle v_\theta \rangle/\partial r$ is negative, the
correlation $\langle v'_r v'_\theta \rangle$ can have a positive value.
On the contrary there is negative $\langle v'_rv'_\theta\rangle$ in region IV with
positive $\partial \langle v_\theta\rangle/\partial \theta$ (see Fig. \ref{mean_deri}b).
\par
The effectiveness of the generation of the positive correlation by the
mean meridional flow can be estimated as follows:
\begin{eqnarray}
\mathcal{M}=- \frac{\partial \langle v_\theta \rangle/\partial r}
  {2\tau\langle\Omega\rangle^2\sin(2\theta)} \sim -\frac{1}{\sin(2\theta)\langle\Omega\rangle}
  \frac{\partial \langle v_\theta\rangle}{\partial r}
  \mathrm{\overline{Ro}},
\end{eqnarray}
where
$\overline{\mathrm{Ro}}=v_\mathrm{rms}/(2\langle \Omega\rangle H_p)$.
When $\mathcal{M}$ is larger than unity, the meridional
flow is effective in generating the correlation 
$\langle v'_rv'_\theta\rangle$.
We note that the mathematical form of
$\mathcal{M}$ indicates that 
it is most difficult to achieve this balance in
mid-latitude due to the factor of $1/\sin(2\theta)$, 
assuming the meridional flow is the same at all latitudes.
Since the positive correlation is found in between $\theta=20$ and
$40\ \mathrm{degrees}$, we estimate $\sin(2\theta)\sim0.5$.
Using the values
$\langle \Omega\rangle/(2\pi)=380\ \mathrm{nHz}$,
$\mathrm{\overline{Ro}}=v_\mathrm{RMS}/(2\langle\Omega\rangle H_p)\sim3$
(which is 
taken from Fig. \ref{rms} at the base of the NSSL) and
$\partial \langle v_\theta \rangle/\partial r\sim -4\times10^{-7}
\mathrm{\ s}^{-1}$
 (around $r=0.95R_\odot$), leads to a value of
$\mathcal{M}$ at the base of NSSL of $1$.
This shows that the generation of the
positive correlation by the mean poleward flow begins to be effective
in the base of the NSSL. When the value 
$\partial \langle v_\theta \rangle/\partial r$ is positive both terms of
the meridional flow and the Coriolis force generate a negative
correlation. This cannot generate the solar-like NSSL even under the
large Rossby number situation (region IV).\par
In the low latitude NSSL (region III), the
positive correlation
$\langle v'_rv'_\theta\rangle$ is mostly generated by the banana cell convection
with some contribution from the poleward meridional flow, where
$\partial \langle v_\theta \rangle/\partial r<0$
(Fig. \ref{vxvy}b). Around the tangential cylinder (white line) the
effect of the
banana cells and the meridional flow is ineffective and the correlation
$\langle v'_rv'_\theta \rangle$ is negative. In the boundary of the
effective and ineffective layer of these mechanisms
i.e. the boundary of the positive and negative
correlation $\langle v'_r v'_\theta\rangle$, the fluid is
accelerated poleward
due to inertial force, which
compensates the Coriolis force in the low latitude NSSL.
The circle in Fig. \ref{vxvy} indicates the boundary area which has
poleward acceleration.\par
 In this study, the equatorward meridional flow in the very near surface
is generated. 
Although the origin of the equatorward
 meridional flow is unknown, this type of feature is seen in the
 previous study \citep{2008ApJ...673..557M}. 
 We find that the equatorward
 meridional flow is generated in the region where the inward directed
 transport stops. This means the angular momentum is deposited
 in this region by the Reynolds stress, which will be transported by the equatorward meridional
 flow. 
 Although in current global calculation we must
 have thick cooling layer ($\sim 4000 \mathrm{\ km}$), in which the
 radial velocity and its radially inward angular momentum transport decrease, the real sun has
 much thinner one ($\sim100\ \mathrm{km}$) in the photosphere. 
The real solar situation
 might not cause a sign change of $d\langle v_\theta \rangle/dr$ in the real sun.
 The distribution of the NSSL especially in the
 low latitude should be confirmed with higher-resolution in the
 future.\par
\section{Summary and Discussion}
We presented a high-resolution, highly stratified numerical
simulation of rotating thermal convection in a spherical shell. We find the self-consistent
generation of a NSSL mostly in high latitudes and analyzed in detail the underlying
angular momentum transport terms and meridional force balance.
\par
 With regard to the angular momentum transport, the maintenance
 mechanism is the same as that suggested by 
 \cite{1975ApJ...199L..71F} and
 \cite{1979ApJ...229.1179G}.
Convection with small rotational influence leads to radially inward transport
of angular momentum.
Since the NSSL deviates
 significantly from the Taylor-Proudman state
($\partial \langle\Omega_1\rangle/\partial z\neq 0$), mechanisms are required
 to balance the Coriolis force which tends to drive
the NSSL towards the Taylor-Proudman state.
These are related to velocity correlations (Reynolds-stresses) within
the meridional plane.
\par
\begin{figure}[htbp]
 \centering
 \includegraphics[width=14cm]{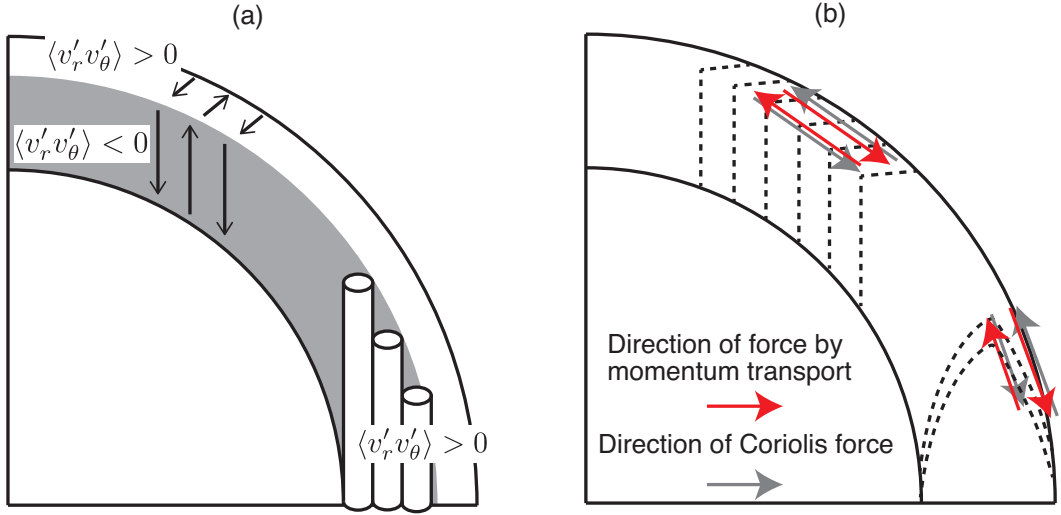}
 \caption{ The summary of our findings in the schematic picture. 
 In this figure we only discuss Reynolds-stress related balances,
 i.e. we do not show the thermal wind balance in the bulk of the
 convection zone.
 The
 panel a shows the distribution of the correlation $\langle
 v'_rv'_\theta\rangle$. The gray area indicates the strong influence of
 the rotation. The panel b shows the force balance on the meridional
 plane. The gray and red arrows show the direction of the Coriolis force
 and the force by the momentum transport. 
Regarding the Coriolis force, the latitudinal component
 is shown.
 The dashed lines are the
 contour line of the angular velocity.
 \label{negativevrvt}} 
\end{figure}
\begin{figure}[htbp]
 \centering
 \includegraphics[width=16cm]{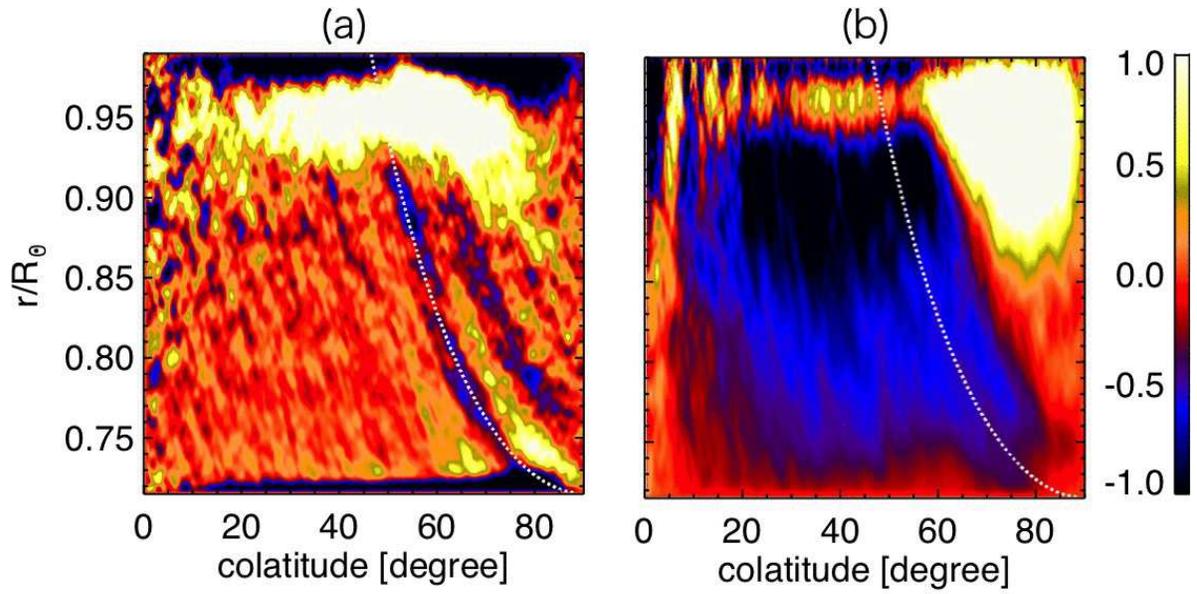}
 \caption{ (a) The quantity 
 $-v_\mathrm{RMS}H_pr\partial (\langle
 v_\theta\rangle/r)/\partial r/3$ is
 shown, which indicates stress by the turbulent viscosity
and
(b) $\langle v'_r v'_\theta\rangle$ for a easy comparison in the unit of 
 in the unit of $10^6\ \mathrm{cm^2\ s^{-2}}$.
 \label{diffusivity}}
\end{figure}

Fig. \ref{negativevrvt}a summarizes the distribution of
the correlations. In the high latitude
 NSSL, a positive correlation $\langle v'_rv'_\theta\rangle$ is generated by the poleward meridional
with negative radial gradient
($\partial \langle v_\theta \rangle/\partial r<0$). 
This can be interpreted as a turbulent viscous stress 
$-\nu_\mathrm{t}r\partial(\langle v_\theta\rangle/r)/\partial r$ in near
 surface layer.
The distribution of estimated turbulent
 viscosity stress is shown in Fig. \ref{diffusivity},
 where the turbulent viscosity
is estimated as $\nu_\mathrm{t}=v_\mathrm{RMS}H_p/3$
 \citep{2012ApJ...751L...9H}. 
\par
Fig. \ref{negativevrvt}b
summarizes the dynamical
balance on the meridional plane.
The poleward meridional flow is generated due to the inward angular
momentum transport. This flow grows until a combination of turbulent
viscous stress and acceleration forces can balance the Coriolis force.
The reason this works in the NSSL is that the radial gradient of the
meridional flow ($d\langle v_\theta \rangle/dr$) is strong and the
RMS velocity is large.
\par
We note that there were some studies that tried to explain differential
rotation through turbulent viscous stresses. This, however, requires,
significantly larger Rossby number, i.e., smaller Taylor number, than
that expected in the solar convection zone 
\citep{1990SoPh..128..243B,1995A&A...299..446K,2006ASPC..354...85W},
which was phrased the ``Taylor-number puzzle'' in the literature.
In this study, this balance
between the Coriolis force and the inertial force
 is well achieved in the high latitude.
In the low latitude, the banana cell generates the positive correlation
 which increases along the radius and accelerates the fluid poleward
 (region III: around the tangential cylinder which is highlighted by
 circle in Fig. \ref{vxvy}).
When the equatorward meridional flow with increasing amplitude
 ($\partial \langle v_\theta\rangle/\partial r>0$) is effective, i.e.,
 the large Rossby number, the correlation $\langle v'_rv'_\theta\rangle$
 becomes negative (region IV). At the layer where this effect begins to occur, the
 fluid is accelerated equatorward. 
Then, the negative correlation becomes zero with approaching the boundary, which then accelerates the fluid
 poleward again.
This complicated transport of momentum governs the meridional force
balance of the NSSL at low latitudes.
\par
In this study, we reduced the solar luminosity to obtain
 the accelerated equator. This reduces the convective velocity and the
 Rossby number. Thus the profile of the NSSL may be also influenced by
 the small Rossby number compared with the actual Sun.
\par
The most important findings in this study are
that the angular momentum is transported radially inward in the NSSL and
that
 the turbulent viscous stress resulting from the radial gradient of
 the latitudinal meridional flow, i.e.,  
$\nu_t r\partial (\langle v_\theta\rangle/r)\partial r$, plays an essential role for the maintenance of
 the NSSL.\par
Our difficulties in obtaining a solar-like profile of the NSSL in very
near surface layer
are possibly
related to the top boundary condition which forces $v_r$ to go to
zero. Observations \citep{2013ApJ...774L..29Z} indicate a poleward flow
with increasing amplitude in radius, which would lead to the proper
positive correlation $\langle v'_rv'_\theta\rangle$ required for a solar-like NSSL.
\acknowledgements
We are grateful to anonymous referee for the helpful comments and
leading us to find the new expression of the RSST.
H. H. is supported by Grant-in-Aid for JSPS Fellows.
The National Center for Atmospheric Research is sponsored by the
National Science Foundation.
The results are obtained by using K computer at the RIKEN
Advanced Institute for Computational Science (Proposal number hp130026
and hp140212). This work was supported in part by MEXT SPIRE and JICFuS.
The authors are grateful to Rachel Howe for giving us the HMI inversion data.
In particular, the authors thank Mark Miesch for educating us about the
role of the gyroscopic pumping and the meridional force balance
and giving us the insightful comments on the manuscript.

\clearpage

\appendix
\section{New expression of RSST}
\label{rsst}
As explained in Paper I, using the original RSST form the momentum and
the total energy are not conserved. In this paper, we adopt the new
expression of the RSST in which these values are mathematically
conserved. 
The conserved value related to the total energy is 
$\rho T s_1 + \rho v^2/2$ and this requires linear
approximation.
Our approach is summarized as: 1. the total density is expressed as 
$\rho=\rho_0 + \tilde{\rho_1}$. 2. using the ordinary linearized
equations of continuity, motion and state with reducing the adiabatic speed of sound and the buoyancy
term 
as:
\begin{eqnarray}
\frac{\partial \tilde{\rho_1}}{\partial t} &=& -\nabla\cdot(\rho{\bf v}),\\
 \rho\frac{\partial {\bf v}}{\partial t} &=&
-\rho({\bf v}\cdot\nabla){\bf v} - \nabla p_1 - \frac{\tilde{\rho_1}}{\xi^2} g
{\bf e_r} + [...],\label{final} \\
p_1 &=& \left(\frac{\partial p}{\partial \rho}\right)_s 
 \frac{\tilde{\rho_1}}{\xi^2}
 +\left(\frac{\partial p}{\partial s}\right)_\rho s_1.\label{eos_s}
\end{eqnarray}
In this idea, we simply reduce the adiabatic speed of sound
$(\partial p/\partial \rho)_s$ by factor of $\xi^2$ with eq. (\ref{eos_s}). 
The balance in the equation of motion makes the perturbation of the
pressure same. This causes the increase of the density perturbation
$\tilde{\rho_1}$ by
factor of $\xi^2$. In order to avoid the increase of the buoyancy, i.e.,
to keep the proper balance between pressure gradient and buoyancy, the
density perturbation for the buoyancy is divided by $\xi^2$ (eq. (\ref{final})).
We tested the validity of this method using a similar way to
\cite{2012A&A...539A..30H}, i.e., a Cartesian box test
problem.
We confirm that with the reduction of the adiabatic speed of sound
scales up the density perturbation by the
factor of $\xi^2$ with remaining the shape of the RMS and
mean density. 
 Tilde
is used, since $\tilde{\rho_1}$ is increased from the ordinary density
perturbation $\rho_1$ by the factor of $\xi^2$.\par
Since the quantity $\tilde{\rho_1}/\xi^2$ remains invariant in leading order
when changing $\xi$, it is more convenient to write $\rho=\rho_0+\xi^2\rho_1$.
Using this expression for the density, we can derive a form of the RSST
that is similar to
\citep{2012A&A...539A..30H},
\begin{eqnarray}
 \frac{\partial \rho}{\partial t}=-\nabla\cdot(\rho {\bf v}) \rightarrow
\frac{\partial \rho_1}{\partial t} = -\frac{1}{\xi^2}\nabla\cdot(\rho
  {\bf v}).
\end{eqnarray}
We note that we use $\rho$ instead of $\rho_0$ in right hand side. In
addition, we also use $\rho$ for the equation of motion and entropy as:
\begin{eqnarray}
 \rho\frac{\partial {\bf v}}{\partial t} &=& -\rho({\bf v}\cdot\nabla){\bf
  v} - \nabla p_1 - \rho_1 g{\bf e_r} + [...],\\
\rho T\frac{\partial s_1}{\partial t} &=& -\rho T ({\bf v\cdot\nabla})s_1 +[...].
\end{eqnarray}
The equation of state is expressed as:
\begin{eqnarray}
 p_1 = \left(\frac{\partial p}{\partial s}\right)_s \rho_1 +
  \left(\frac{\partial p}{\partial \rho}\right)_\rho s_1.
\end{eqnarray}
Then the variable $\rho$ is conserved mathematically.
We again note that in this discussion
$\rho=\rho_0+\xi^2\rho_1$.
In addition, the expressions
\begin{eqnarray}
 \rho\frac{\partial {\bf v}}{\partial t} + \rho({\bf v}\cdot\nabla){\bf
  v},
\end{eqnarray}
and
\begin{eqnarray}
 \frac{\partial }{\partial t}(\rho {\bf v}) + \nabla\cdot(\rho {\bf vv}),
\end{eqnarray}
are identical. This means that the angular momentum $\rho\mathcal{L}$ is
conserved with this form mathematically.
\par
Next, we derive the conservation of total energy under the linear
approximation, i.e., ignore the second order term.
From the hydrostatic equilibrium, the relation
\begin{eqnarray}
 g &=& -\frac{1}{\rho_0}\frac{dp_0}{dr} \nonumber\\
   &=& -\frac{1}{\rho_0 v_r}\frac{Dp_0}{Dt},
\end{eqnarray}
is obtained, where $D/Dt=\partial/\partial t +{\bf v}\cdot\nabla$ is
the Lagrangian derivative.
The equation of the kinetic energy is written as:
\begin{eqnarray}
\rho\frac{D}{Dt}\left(\frac{1}{2}v^2\right) +({\bf v}\cdot\nabla)p_1 +
 v_r\rho_1 g &=& \rho\frac{D}{Dt}\left(\frac{1}{2}v^2\right)\nonumber\\
 &&+\nabla\cdot({\bf v}p_1) 
 +\frac{p_1}{\rho}\frac{{D\rho}}{Dt}
 -\frac{\rho_1}{\rho_0}\frac{Dp_0}{Dt}\nonumber\\
&=&0.
\end{eqnarray}
Our background temperature gradient is adiabatic:
\begin{eqnarray}
 s_1\frac{DT}{Dt} \sim s_1\frac{DT_0}{Dt} &=& 
  s_1\left(\frac{\partial
   T}{\partial
   \rho}\right)_s\frac{D\rho_0}{Dt}\\
 &=& 
  s_1\left(\frac{\partial
   T}{\partial
   p}\right)_s\frac{Dp_0}{Dt}.
\end{eqnarray}
Then the equation of entropy ($\rho TDs_1/Dt=Q$, where $Q$ includes
radiative diffusion and surface cooling), is transformed as:
\begin{eqnarray}
 \rho\frac{D}{Dt}\left(Ts_1\right) - \rho s_1\frac{DT}{Dt}
  &\sim&  \rho\frac{D}{Dt}\left(Ts_1\right) - \rho
\left[
\left(\frac{\partial s}{\partial \rho}\right)_p \rho_1 +
\left(\frac{\partial s}{\partial p}\right)_\rho p_1
\right]\frac{DT_0}{Dt}\nonumber\\
&=& \rho\frac{D}{Dt}\left(Ts_1\right)
-\rho\left[
\left(\frac{\partial s}{\partial \rho}\right)_p
\left(\frac{\partial T}{\partial p}\right)_s
\rho_1 \frac{Dp_0}{Dt}
+
\left(\frac{\partial s}{\partial p}\right)_\rho
\left(\frac{\partial T}{\partial \rho}\right)_s
p_1 \frac{D\rho_0}{Dt}
\right]\nonumber\\
&=& \rho\frac{D}{Dt}\left(Ts_1\right)
-\rho\left[
\left(-\frac{c_\mathrm{p}}{\beta\rho_0T_0}\right)
\left(\frac{\beta T_0}{c_\mathrm{p}\rho_0}\right)
\rho_1 \frac{Dp_0}{Dt}
+
\left(\frac{\kappa_\mathrm{T}c_\mathrm{v}}{\beta T_0}\right)
\left(\frac{\beta T_0}{c_\mathrm{v}\kappa_\mathrm{T}\rho_0^2}\right)
p_1 \frac{D\rho_0}{Dt}
\right]\nonumber\\
&=& \rho\frac{D}{Dt}\left(Ts_1\right)
-\rho\left(
-\frac{\rho_1}{\rho^2_0}\frac{Dp_0}{Dt}+\frac{p_1}{\rho_0^2}\frac{D\rho_0}{Dt}
\right) = Q,
\end{eqnarray}
where $\beta$ and $\kappa_\mathrm{T}$ are the coefficient of thermal
expansion and the coefficient of isothermal compressibility,
respectively \citep{2014ApJ...786...24H,1984oup..book.....M}. Thus the
equation of the total energy is expressed with using the linear approximation
\begin{eqnarray}
 \rho\frac{D}{Dt}\left(Ts_1+\frac{1}{2}v^2\right) +\nabla\cdot({\bf
  v}p_1) = Q.
\end{eqnarray}
The value $\rho Ts_1+\rho v^2/2$ is conserved. 
We note the deviation is mainly caused by the value
$\xi^2\rho_1/\rho_0$.
Using the equation of state for the perfect gas, the value is
transformed as
$\rho Ts_1\sim \rho c_\mathrm{v} T_1 - p_0 \rho_1/\rho_0$, which means
the internal energy and contribution of the buoyancy. We note that using
the anelastic approximation $(0=\nabla\cdot(\rho_0{\bf v}))$, the value
$\rho_0T_0s_1+\rho_0 v^2/2$ is conserved without any linear
approximation. 
The above derivation assumes the adiabatic background stratification.
Some additional terms and assumptions would be required when we have
non-adiabatic background stratification
\par
\section{Artificial viscosity}
\label{art}
The same artificial viscosity as MuRAM code \citep{2014ApJ...789..132R} is added on
all the variables as:
\begin{eqnarray}
&& \frac{\partial}{\partial t}(\rho_1 \xi^2) = -\nabla\cdot{\bf F_\rho},\\
&&\rho \frac{\partial v_r}{\partial t} = -\nabla\cdot{\bf F_{v_r}},\\
&&\rho \frac{\partial v_\theta}{\partial t} = -\nabla\cdot{\bf F_{v_\theta}},\\
&&\rho \frac{\partial v_\phi}{\partial t} = -\nabla\cdot{\bf F_{v_\phi}},\\
&&\rho T\frac{\partial s_1}{\partial t} = -\nabla\cdot{\bf F_{s}}.
\end{eqnarray}
  \begin{eqnarray}
   F_{i+1/2}=-\frac{1}{2}c_{i+1/2}\phi_{i+1/2}\left(u_\mathrm{r}-u_\mathrm{l},u_{i+1}-u_i\right)
    \left(u_\mathrm{r}-u_\mathrm{l}\right),\\
    \phi =\left\{ 
	   \begin{array}{cc}
	    \displaystyle{
	     \mathrm{max}\left[0,1+h\left(\frac{u_r-u_l}{u_{i+1}-u_i}-1\right)\right]
	     }
&
	     \mathrm{for}\
	     (u_\mathrm{r}-u_\mathrm{l})\cdot(u_{i+1}-u_i)>0,\\
	    0 &
	     \mathrm{for}\
	     (u_\mathrm{r}-u_\mathrm{l})\cdot(u_{i+1}-u_i)\le0,
	    \label{append:diffusive}
	   \end{array}
       \right.
  \end{eqnarray}
  where $c_{i+1/2}=0.3c_\mathrm{s}+v $ is the characteristic
  velocity which is the sum of the speed of sound ($c_\mathrm{s}$) and
  fluid velocity ($v$).
  To decrease the effect of viscosity, a multiplier 0.3 is used and
  h=0.75 is adopted.
  In the code, the physical variables $u_i$ are defined at the center of
  the cell. To calculate the diffusive flux, the variables
  $u_\mathrm{r}$ and $u_\mathrm{l}$ at a boundary of the cells are
  defined as:
  \begin{eqnarray}
&& u_\mathrm{l} = u_i + \frac{1}{2}\Delta u_i,\\
   && u_\mathrm{r} = u_{i+1} - \frac{1}{2}\Delta u_{i+1},
  \end{eqnarray}
  where the tilt of the variable $(\Delta u_i)$ is defined as:
  \begin{eqnarray}
   \Delta u_i = \mathrm{minimod}
    \left(
     \epsilon(u_{i+1}-u_{i}),\frac{u_{i+1}-u_{i-1}}{2},\epsilon(u_i-u_{i-1})
	    \right),
  \end{eqnarray}
  where $\epsilon$ is the factor for the minimod function
  ($1<\epsilon<2$), in this study $\epsilon=1.4$ is used.
  To conserve total energy, the heat
  from the dissipated kinetic energy is treated accordingly.
  The heat caused by the artificial
  viscosity is estimated and added in the equation of entropy as
  \begin{eqnarray}
   \rho T \frac{\partial s_1}{\partial t} = 
-({\bf F}_r\cdot\nabla) v_r 
-({\bf F}_\theta\cdot\nabla) v_\theta
-({\bf F}_\phi\cdot\nabla) v_\phi.
  \end{eqnarray}
\section{Dynamical balance on the meridional plane}\label{twb}
\begin{figure}[htbp]
 \centering
 \includegraphics[width=16cm]{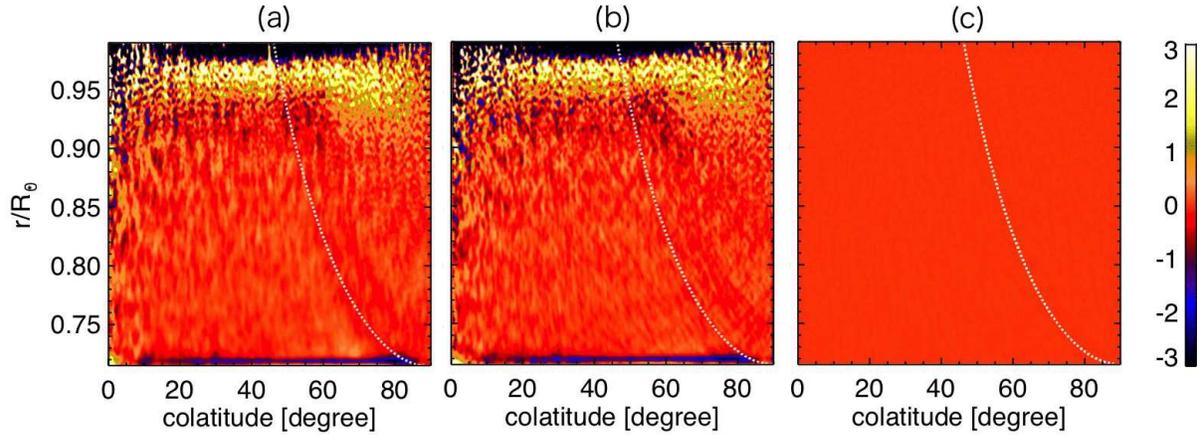}
 \caption{ The values (a) $\nabla\times(\langle{\bf v}\times{\bf \omega}\rangle)$, 
 (b) $\mathcal{C}=-\nabla\times[\nabla\cdot(\rho_0\langle{\bf vv}\rangle)/\rho_0]$, and
 (c) $\nabla\times[\langle{\bf v}\nabla\cdot(\rho{\bf v})/\rho\rangle]$
are shown in the unit
 of $10^{-12}\ \mathrm{s^{-2}}$
 are shown on the meridional plane. The white lines show the location of
 the tangential cylinder.
 \label{compare}} 
\end{figure}

In the appendix, we derive the equations for the
dynamical balance on the meridional plane.
We start with the hydrodynamic equation with the Coriolis
force used in this paper (eq. (\ref{eqmo})) 
\begin{eqnarray}
\frac{\partial {\bf v}}{\partial t} 
= 
-({\bf v}\cdot\nabla){\bf v}
-\frac{\nabla p_1 + \rho_1g{\bf e_r}}{\rho}
+2{\bf v}\times{\bf \Omega_0}+{\bf G},
\label{oeqmo}
\end{eqnarray}
where final term shows the artificial viscosity
${\bf G}=-(\nabla\cdot {\bf F_{v_r}}){\bf e_r} -(\nabla\cdot {\bf
F_{v_\theta}}){\bf e_\theta}
 -(\nabla\cdot {\bf F_{v_\phi}}){\bf e_\phi}$ (see Appendix \ref{art}).
The curl of the first term in the right hand side
of eq. (\ref{oeqmo}) is transformed as
$\nabla\times ({\bf v}\times{\bf \omega})$
with using the vector formula
\begin{eqnarray}
 ({\bf v}\cdot\nabla){\bf v} = \nabla\left(\frac{v^2}{2}\right) 
  - {\bf v}\times(\nabla\times{\bf v}).
\end{eqnarray}
Although in this paper, we directly take the curl of the second term, it
is useful to show the zonal component of the curl of the second term in
the right hand side of eq. (\ref{oeqmo}) with using $\rho_0$ instead of
$\rho$ as:
\begin{eqnarray}
\left[
\nabla\times 
\left(
 -\frac{\nabla p_1 + \rho_1g{\bf e_r}}{\rho_0}
\right)\right]_\phi
&=& \frac{1}{\rho_0^2r}\frac{d \rho_0}{d r}\frac{\partial p_1}{\partial
\theta}
+\frac{g}{\rho_0r}\frac{\partial \rho_1}{\partial \theta}\nonumber\\
&=&
-\frac{g}{\rho_0r}\left[
\left(\frac{\partial \rho}{\partial p}\right)_s
\frac{\partial p_1}{\partial \theta}
-\frac{\partial \rho_1}{\partial \theta}
\right]\nonumber\\
&=&
\frac{g}{\rho_0r}\left(\frac{\partial \rho}{\partial s}\right)_p
\frac{\partial s_1}{\partial \theta},
\end{eqnarray}
We note that for the perfect gas the value 
$(\partial \rho/\partial s)_p=-\rho_0/c_\mathrm{p}$, where
$c_\mathrm{p}$ is the heat capacity at constant volume.
Next, the zonal component of the curl of the third term in the left hand
size of eq. (\ref{oeqmo}) is transformed as:
\begin{eqnarray}
\mathcal{T}_0 = [\nabla\times (2{\bf v}\times {\bf \Omega}_0)]_\phi 
   &=& [2({\bf \Omega}_0\cdot\nabla){\bf v_r}-2({\bf v}\cdot\nabla){\bf
   \Omega}_0]_\phi\nonumber\\
 &=& 2({\bf \Omega_0}\cdot\nabla) v_\phi =
  2r\sin\theta\Omega_0\frac{\partial \Omega_1}{\partial z},
\end{eqnarray}
where $\Omega_1=v_\phi/(r\sin\theta)$. In the transformation, the
formulation $\nabla\cdot{\bf \Omega_0}=0$, ${\bf \Omega_0}\cdot{\bf
e_\phi}=0$, are used.
Fig. \ref{compare} shows that the values 
$[\nabla\times(\langle{\bf v\times}{\bf \omega}\rangle)]_\phi$ and
$-(\nabla\times[\nabla\cdot(\rho_0\langle{\bf vv}\rangle)/\rho_0])_\phi$ are almost
equivalent and $\langle\nabla\times({\bf v}/\rho\nabla\cdot(\rho{\bf
v}))\rangle$ is fairly small compared with the other terms. 
Thus it is valid to use $-(\nabla\times[\nabla\cdot(\rho_0\langle{\bf
vv}\rangle)/\rho_0])_\phi$ in stead of $[\nabla\times(\langle{\bf v\times}{\bf \omega}\rangle)]_\phi$.
Then we define the momentum flux on the meridional plane as:
\begin{eqnarray}
&& \langle F_{ij}\rangle = \tilde{F_{ij}} + F'_{ij}\\
&& \tilde{F_{ij}} = \rho_0\langle v_i\rangle \langle v_j\rangle,\\
&& F'_{ij} = \rho_0\langle v'_i v'_j\rangle,
\end{eqnarray}
where $i$ and $j$ correspond to $r$, $\theta$, and $\phi$. 
For this definition, we divide the velocity as
$v_i=\langle v_i\rangle+v'_i$.
Then the
divergence of the fluxes are divided to several terms as:
\begin{eqnarray}
 && {\bf D} = -\frac{1}{\rho_0}\nabla\cdot {\bf F} =
D_r{\bf e_r} + D_\theta{\bf e_\theta},\\
 && D_r = D_{r\mathrm{(d)}} + D_{r\mathrm{(n)}},\\
 && D_\theta = D_{\theta\mathrm{(d)}} + D_{\theta\mathrm{(n)}},\\
 && D_{r(\mathrm{d})} = -\frac{1}{\rho_0}
  \left[
   \frac{1}{r^2}\frac{\partial }{\partial r}(r^2 F_{rr})
   -\frac{F_{\theta\theta}}{r}
  \right],\\
 && D_{r(\mathrm{n})} = -\frac{1}{\rho_0r\sin\theta}\frac{\partial}{\partial
  \theta} (\sin\theta F_{\theta r}),\\
 && D_{\theta(\mathrm{d})} = 
  -\frac{1}{\rho_0}
   \frac{1}{r\sin\theta}\frac{\partial}{\partial
   \theta} (\sin\theta F_{\theta \theta}),\\
 && D_{\theta(\mathrm{n})} = -\frac{1}{\rho_0}
  \left[
   \frac{1}{r^2}\frac{\partial }{\partial r}(r^2 F_{r\theta})
   +\frac{F_{\theta r}}{r}
	  \right],\\
&& D_{r\phi} = \frac{F_{\phi\phi}}{r\rho_0},\\
&& D_{\theta\phi} = \cot \theta\frac{F_{\phi\phi}}{r\rho_0},
\end{eqnarray}
We use the notation of $\tilde{D}=D(\tilde{F})$ and $D'=D(F')$. Then the
zonal component of the curl of the ${\bf \langle D\rangle}$ is also divided to several
terms as:
\begin{eqnarray}
&& \mathcal{C} = (\nabla\times \langle{\bf D}\rangle)_\phi 
   = \mathcal{C}_r + \mathcal{C}_\theta + \mathcal{C}_\mathrm{d}.
\end{eqnarray}
Then each term is divided 
$\mathcal{C}_i = \tilde{\mathcal{C}}_i + \mathcal{C}'_i$, where $i$
corresponds to $r$, $\theta$, $\mathrm{d}$. The terms are
\begin{eqnarray}
\tilde{\mathcal{C}}_r &=& -\frac{1}{r}\frac{\partial \tilde{D}_{r(\mathrm{n})}}
{\partial \theta},\ 
\mathcal{C}'_r = -\frac{1}{r}\frac{\partial D'_{r(\mathrm{n})}}
{\partial \theta}, 
\\
\tilde{\mathcal{C}}_\theta &=& \frac{1}{r}\frac{\partial }{\partial r}\left(r
 \tilde{D}_{\theta(\mathrm{n})}\right),\
\mathcal{C}'_\theta = \frac{1}{r}\frac{\partial }{\partial r}\left(r
 D'_{\theta(\mathrm{n})}\right),
\\
\tilde{\mathcal{C}}_\mathrm{d} &=&
 \frac{1}{r}\frac{\partial }{\partial r}(r
 \tilde{D}_{\theta(\mathrm{d})})
-\frac{1}{r}\frac{\partial \tilde{D}_{r(\mathrm{d})}}{\partial \theta},\\
\mathcal{C'}_\mathrm{d} &=&
 \frac{1}{r}\frac{\partial }{\partial r}
\left[r
 (D'_{\theta(\mathrm{d})}+D'_{\theta\phi})
\right]
-\frac{1}{r}\frac{\partial}{\partial \theta}
\left(D'_{r\mathrm{(d)}}+D'_{r\phi}\right),\\
\mathcal{T}_1 &=&
 \frac{1}{r}\frac{\partial }{\partial r}(r
 \tilde{D}_{\theta\phi})
-\frac{1}{r}\frac{\partial \tilde{D}_{r\phi}}{\partial \theta},\nonumber\\
 &=&\frac{\sin\theta\cos\theta}{r}\frac{\partial}{\partial r}
  \left(r^2\langle\Omega_1\rangle^2\right) 
  - \frac{\partial}{\partial \theta}\left(\sin^2\theta \langle
				     \Omega_1\rangle^2\right)
  \nonumber\\
&=& r\sin\theta \frac{\partial \langle\Omega_1\rangle^2}{\partial z}.
\end{eqnarray}
We note that the contribution of the mean differential rotation 
$\langle \Omega_1\rangle$ is separated with using the term
$\mathcal{T}_1$.
Then the eq. (\ref{oeqmo}) is averaged in time and zonal direction.
The equation of the balance is
obtained as:
\begin{eqnarray}
 \mathcal{W}-\mathcal{T} = \mathcal{B}+\mathcal{C}+\mathcal{V},\\
\end{eqnarray}
where
\begin{eqnarray}
 \mathcal{W} &=& \frac{\partial \langle \omega_\phi \rangle}{\partial t}, \\
 \mathcal{T} &=& \mathcal{T}_0 + \mathcal{T}_1,\nonumber\\
 &=&r\sin\theta \frac{\partial \langle\Omega\rangle^2}{\partial z},\\
 \mathcal{B} &=&
  -\left[\nabla\times\left(\frac{\nabla p_1 + \rho_1 g {\bf e_r}}{\rho}\right)\right]_\phi,\\
 \mathcal{V} &=& [\langle\nabla\times{\bf G}\rangle]_\phi.
\end{eqnarray}

\clearpage

\clearpage

\begin{figure}[htbp]
 \centering
 \includegraphics[width=16cm]{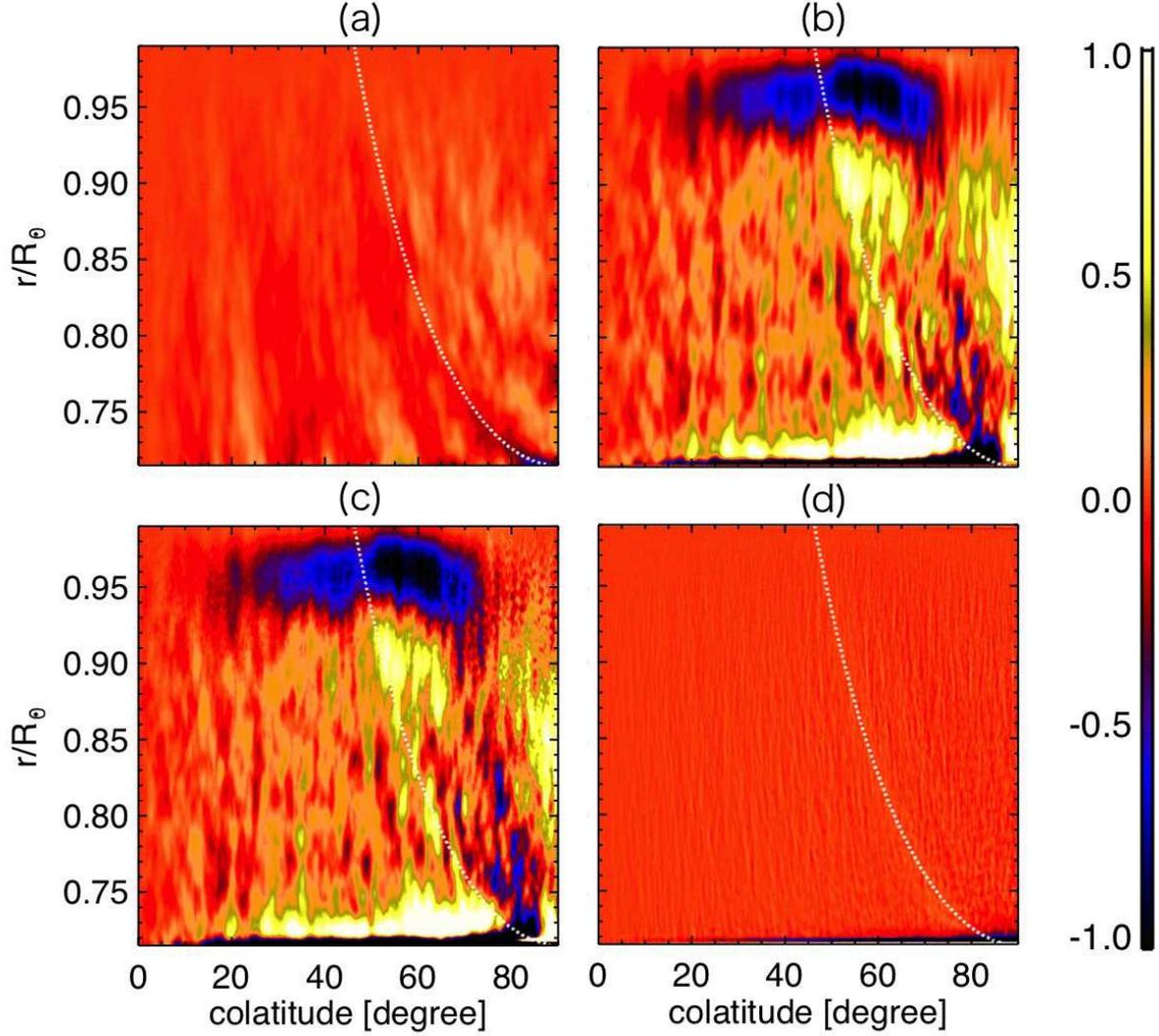}
 \caption{
 The values
 (a) $\rho_0\partial \langle \mathcal{L}\rangle/\partial t$,
 (b) $\rho_0\langle{\bf v_\mathrm{m}}\rangle\cdot\langle \mathcal{L}\rangle$, 
(c) $-\rho_0\langle ({\bf v'_\mathrm{m}}\cdot\nabla)\mathcal{L}'\rangle$
and (d) $-\rho_0r\sin\theta\left\langle \nabla\cdot {\bf
 F_{v_\phi}}/\rho \right\rangle$ in the unit of 
$10^6\ \mathrm{g\ cm^{-1}\ s^{-2}}$ are shown on the meridional
 plane. The white lines show the location of the tangential cylinder.
 \label{ang_balance}} 
\end{figure}

\begin{figure}[htbp]
 \centering
 \includegraphics[width=16cm]{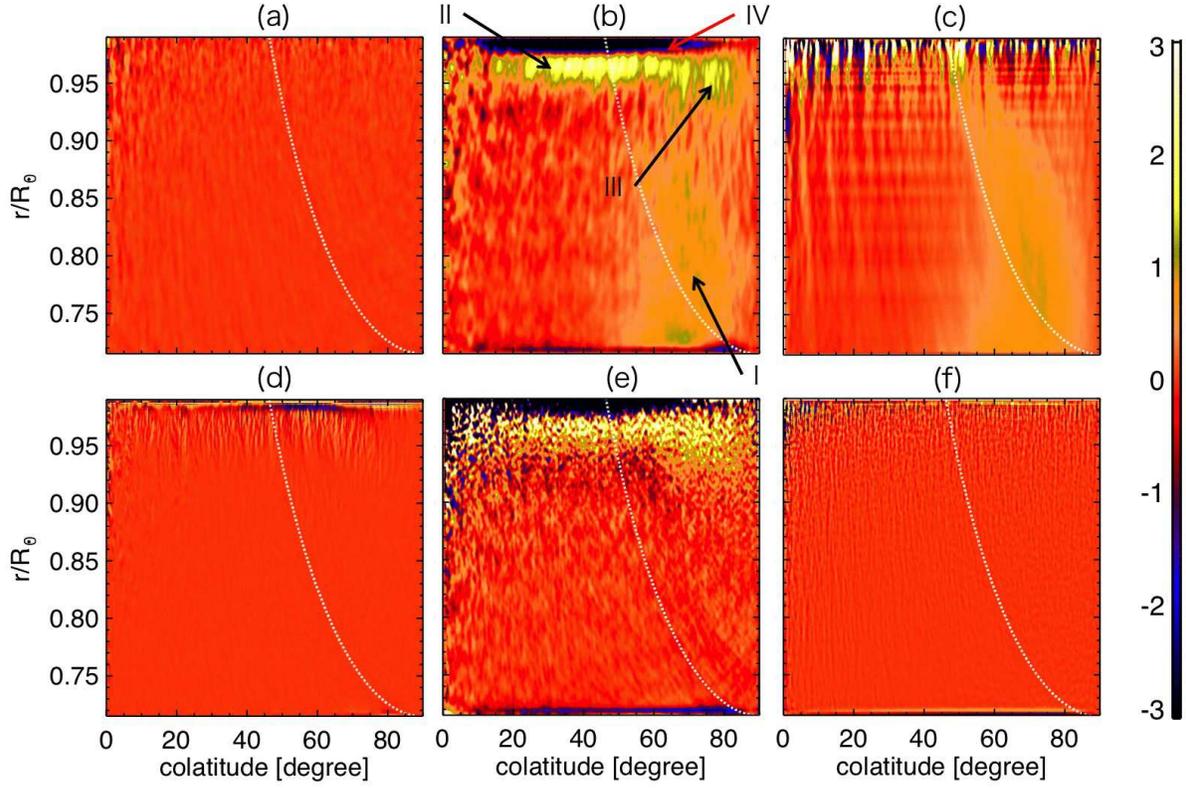}
 \caption{
The values (a) $\mathcal{W}$, (b) $\mathcal{-T}$, (c) $\mathcal{B}$, 
 (d) $\tilde{\mathcal{C}}$, (e)
 $\mathcal{C}'$, and (f) $\mathcal{V}$
 are shown in the unit of $10^{-12}\ \mathrm{s^{-2}}$
are shown on the meridional plane. The white lines show the location of
 the tangential cylinder.
 The indicated regions I-IV refer to different balances achieved:  I:
 $-\mathcal{T}=B$, II, III, and IV: $-\mathcal{T} =\mathcal{C}'$.
 \label{term}} 
\end{figure}

\end{document}